\documentclass[12pt,onecolumn,a4paper,draftcls]{IEEEtran}
\usepackage{mathrsfs}
\usepackage{}
\usepackage{amssymb}
\usepackage{amsfonts}
\usepackage{epstopdf}
\usepackage{graphicx}
\usepackage{stmaryrd}
\usepackage{amssymb,amsmath}
\usepackage{multirow,url}
\usepackage{color,cite}
\usepackage{hyperref}

\ifodd 1
\newcommand{\com}[1]{\textbf{\color{red} (COMMENT: #1)}} 
\newcommand{\comg}[1]{\textbf{\color{green} (COMMENT: #1)}}
\newcommand{\response}[1]{\textbf{\color{magenta} (RESPONSE: #1)}} 
\else

\newcommand{\com}[1]{}
\newcommand{\comg}[1]{}
\newcommand{\response}[1]{}
\fi
\ifodd 0
\newcommand{\referred}[1]{\textcolor{red}{RefPaper: #1}} 
\else
\newcommand{\referred}[1]{}
\fi

\ifodd 1
\newcommand{\changeblue}[1]{\textcolor{blue}{Modified: #1}} 
\else
\newcommand{\changeblue}[1]{}
\fi

\newtheorem{definition}{Def\/inition}

\ifCLASSINFOpdf
\else
\fi
\begin{document}
%
\title{Asynchronous Physical-layer Network Coding}

%

%
%

\author{Lu~Lu,~\IEEEmembership{Student Member,~IEEE,}
       and~Soung~Chang~Liew,~\IEEEmembership{Senior Member,~IEEE}
\thanks{L. Lu and S. C. Liew are with the Department
of Information Engineering, The Chinese University of Hong Kong, Hong Kong.
e-mails: \{ll007, soung\}@ie.cuhk.edu.hk}
\thanks{An early version of this work addressing only unchannel-coded PNC appeared in \emph{Proc. IEEE ICC} 2011 \cite{UPNCICC11}.}
}






\maketitle

\begin{abstract}
A key issue in physical-layer network coding (PNC) is how to deal with the asynchrony between signals transmitted by multiple transmitters. That is, symbols transmitted by different transmitters could arrive at the receiver with symbol misalignment as well as relative carrier-phase offset. A second important issue is how to integrate channel coding with PNC to achieve reliable communication. This paper investigates these two issues and makes the following contributions:
1) We propose and investigate a general framework for decoding at the receiver based on belief propagation (BP). The framework can effectively deal with symbol and phase asynchronies while incorporating channel coding at the same time.
2) For unchannel-coded PNC, we show that for BPSK and QPSK modulations, our BP method can significantly reduce the asynchrony penalties compared with prior methods.
3) For unchannel-coded PNC, with half symbol offset between the transmitters, our BP method can drastically reduce the performance penalty due to phase asynchrony, from more than 6 dB to no more than 1 dB.
4) For channel-coded PNC, with our BP method, both symbol and phase asynchronies actually improve the system performance compared with the perfectly synchronous case. Furthermore, the performance spread due to different combinations of symbol and phase offsets between the transmitters in channel-coded PNC is only around 1 dB.
The implication of 3) is that if we could control the symbol arrival times at the receiver, it would be advantageous to deliberately introduce a half symbol offset in unchannel-coded PNC. The implication of 4) is that when channel coding is used, symbol and phase asynchronies are not major performance concerns in PNC.

\end{abstract}


%
\IEEEpeerreviewmaketitle

\thispagestyle{empty}
\newpage
\setcounter{page}{1}

\section{Introduction}
%
%

Physical-layer network coding (PNC), first proposed in \cite{PNC06}, is a subfield of network coding \cite{AhlswIT00} that is attracting much attention recently. The simplest system in which PNC can be applied is the two-way relay channel (TWRC), in which two end nodes exchange information with the help of a relay node in the middle, as illustrated in Fig. \ref{fig:system}(a). This paper focuses on TWRC. Compared with the conventional relay system, PNC doubles the throughput of TWRC by reducing the needed time slots for the exchange of one packet from four to two.

In PNC, in the first time slot, the two end nodes send signals simultaneously to the relay; in the second phase, the relay processes the superimposed signals of the simultaneous packets and maps them to a network-coded packet for broadcast back to the end nodes.

A key issue in PNC is how to deal with the asynchronies between the signals transmitted simultaneously by the two end nodes. That is, symbols transmitted by the two end nodes could arrive at the receiver with symbol misalignment as well as relative carrier-phase offset.

Many previous works (e.g., \cite{PNC06, SyncPNC06, HaoMilcom07}) found that symbol misalignment and carrier-phase offset will result in appreciable performance penalties. For BPSK modulation, \cite{PNC06} showed that the BER performance penalties due to the carrier-phase offset and symbol offset are both 3 dB in the worst case. For QPSK modulation, the penalty can be as large as 6 dB in the worst case when the carrier-phase offset is $\pi/4$ \cite{HaoMilcom07}. These results are for unchannel-coded PNC.

These earlier investigations led to a common belief that near-perfect symbol and carrier-phase synchronizations are important for good performance in PNC. This paper shows that this is not exactly true, and that asynchronous PNC can have good performance when appropriate methods are applied.

The study of BPSK unchannel-coded PNC in \cite{PNC06} and \cite{SyncPNC06}, for example, made use of suboptimal decoding methods at the relay for asynchronous PNC. Furthermore, the joint effect of symbol and phase asynchronies was not investigated. In this paper, we propose an optimal maximum-likelihood (ML) decoding method that makes use of a belief propagation (BP) algorithm. Our method addresses symbol and phase asynchronies jointly within one framework. We find that our method can reduce the worst-case BER performance penalty of 3 dB in \cite{PNC06} and \cite{SyncPNC06} to less than 0.5 dB.

In QPSK unchannel-coded PNC, the penalty is larger than 6 dB \cite{HaoMilcom07} only when the symbols are perfectly aligned and when the phase offset is $\pi/4$ (benchmarked against the perfectly synchronous case in which there are no symbol and phase offsets). We find that using our method, when there is a half symbol misalignment, the penalty is reduced to less than 1 dB. Additionally, an interesting result is that with half symbol misalignment, the spread of penalties under various carrier-phase offsets is no more than 0.5 dB. This means that symbol misalignment has the effect of desensitizing the performance of the system to carrier-phase offset.

For reliable communication, channel coding is often applied. Therefore, another important issue is how to incorporate channel coding into PNC. We extend our BP method so that it can incorporate channel decoding and deal with asynchrony at the same time. An interesting result when channel coding is adopted is that with an appropriate BP algorithm, instead of asynchrony penalty, we have asynchrony reward. In particular, both symbol misalignment and phase offset improve BER performance when channel coding is used. In addition, the performance spread arising from all combinations of symbol and phase offsets is only around 1 dB. This suggests that when channel coding is used, symbol and phase asynchronies are not major performance concerns.

The remainder of this paper is organized as follows: Section \ref{Sec:RelatedWork} overviews related work. Section \ref{Sec:SystemModel} introduces the system model of this paper. Section \ref{Sec:UPNC} presents our BP ML decoding method for asynchronous unchannel-coded PNC. Section \ref{Sec:CPNC} extends the method for channel-coded PNC. Numerical results are given in the subsections of \ref{Sec:UPNC3} and \ref{Sec:CPNC4}, respectively. Finally, Section \ref{Sec:Conclusion} concludes this paper.

\section{Related Work} \label{Sec:RelatedWork}
Prior work falls into two main categories: schemes for unchannel-coded PNC, and schemes for channel-coded PNC. Our paper, however, is the first paper that studies the joint effect of symbol and phase asynchronies in both unchannel-coded and channel-coded PNC.



\subsection{Classification} \label{Classification}
Table \ref{tab:Class} shows the four possible cases for PNC systems. In Table \ref{tab:Class}, $\Delta \in [0, T)$ is the relative symbol offset between the two end nodes $A$ and $B$, where $T$ is the symbol duration; and $\phi \in [0, 2\pi)$ is the relative phase offset between the RF carriers of the two end nodes. Case 1 is the perfectly synchronized case studied in \referred{PNC06, PopICC07} \cite{PNC06, PopICC07}; Case 2 is the symbol-asynchronous case studied in \referred{SyncPNC06}\cite{SyncPNC06}; and Case 3 is the phase-asynchronous case studied in \referred{SyncPNC06, ShengliGlobecom08, KoikeJSAC09, WubbenGlobecom10}\cite{SyncPNC06, ShengliGlobecom08, KoikeJSAC09, WubbenGlobecom10}. To our best knowledge, the asynchronous-symbol asynchronous-phase Case 4 has not been studied before. This paper proposes a general scheme to tackle all four cases under one framework.

\subsection{Unchannel-coded PNC}

Although asynchronous unchannel-coded PNC has been studied previously, only suboptimal decoding algorithms were considered. Refs. \cite{PNC06} and \cite{SyncPNC06} argued that the largest BER performance penalty is 3 dB for BPSK modulation (for both phase and symbol asynchronies). However, this conclusion is based on suboptimal decoding.

Ref. \cite{HaoMilcom07} mentioned without proof that there is a maximum 6 dB BER performance penalty for QPSK modulation when $\Delta = 0$ and $\phi \ne 0$. To the best of our knowledge, no quantitative results and concrete explanation have been given for the general $\Delta$ and $\phi$ case.

Ref. \cite{KoikeJSAC09} investigated systems in which symbols are aligned but phases are not. It uses QPSK for uplinks, but a higher order constellation map (e.g., 5QAM) for downlinks when the uplink phase offset is not favorable for QPSK downlink. That is, it varies the mode of PNC mapping depending on the phase asynchrony. In this paper, we assume the simpler system in which both the uplink and downlink use the same modulation, either BPSK or QPSK.

\subsection{Channel-coded PNC}

For channel-coded PNC, an important issue is how to integrate the channel decoding operation and the network coding operation at the relay. Ref. \referred{ShengliJSAC09}\cite{ShengliJSAC09} presented a scheme that works well for synchronous channel-coded PNC. The scheme is not amenable to extension for asynchronous channel-coded PNC.

Ref. \referred{WubbenGlobecom10}\cite{WubbenGlobecom10} proposed a method for phase-asynchronous channel-coded PNC, assuming the use of Low-Density Parity-Check (LDPC) code. Different from the scheme in \referred{WubbenGlobecom10}\cite{WubbenGlobecom10}, our method deals with both the phase and symbol asynchronies.

Refs. \referred{ZorziSPAWC09}\cite{ZorziSPAWC09} and \referred{FPNC11}\cite{FPNC11} investigated OFDM PNC. With OFDM, the symbol offset in the time domain is translated into different phase offsets in different subcarriers in the frequency domain. Since different subcarriers experience different phase offsets, there is an averaging effect as far as performance is concerned, and the system performance is not at the mercy of the worst-case phase asynchrony. The channel-decoding and network-coding process in \referred{ZorziSPAWC09, FPNC11}\cite{ZorziSPAWC09} and \cite{FPNC11}, however, are performed in a disjoint manner (using an XOR-CD decoder that will be described in Section \ref{Sec:CPNC3}). By contrast, the joint channel-decoding and network-coding scheme (Jt-CNC) proposed in this paper can yield much better performance (to be presented in Section \ref{Sec:CPNC2}).

In this paper, we use the Repeat-Accumulate (RA) channel code to explain the principles of Jt-CNC and XOR-CD, as well as in the numerical studies. We believe that the general conclusions will be the same if the LDPC code is used instead. The application of the convolutional code in PNC has been studied previously for symbol-synchronous PNC \referred{ToTWireless10}\cite{ToTWireless10}. However, BP decoding was not used. The use of BP decoding for convolutional-coded PNC is an interesting area for further work because the convolutional code has lower complexity than the RA and LDPC codes.

The use of BP has also been proposed in a number of places in the literature in the context of multi-user detection \referred{BoutrosIT02}\cite{BoutrosIT02} and joint detection and decoding in the presence of phase noise and frequency offset \referred{BarbieriTCom07}\cite{BarbieriTCom07}. In \referred{ZhuTWireless09}\cite{ZhuTWireless09}, BP over a factor graph that describes the joint probability law of all unknowns and observations, as in the framework used in our paper here, is used to decode one of two users in the presence of Gauss-Markov (non-block) fading. This work is along the line of collision resolution \referred{CRESM}\cite{CRESM} rather than PNC mapping. In this paper, we assume that the channels can be perfectly estimated, leaving out the detailed estimation procedure. Ref. \referred{ZhuTWireless09}\cite{ZhuTWireless09} provides a nice way to integrate the problem of channel estimation and detection using BP. The application of the technique in PNC is an interesting area for further work.

\section{System Model}\label{Sec:SystemModel}

\begin{figure}[h]
\centering
\includegraphics[width=0.8\textwidth]{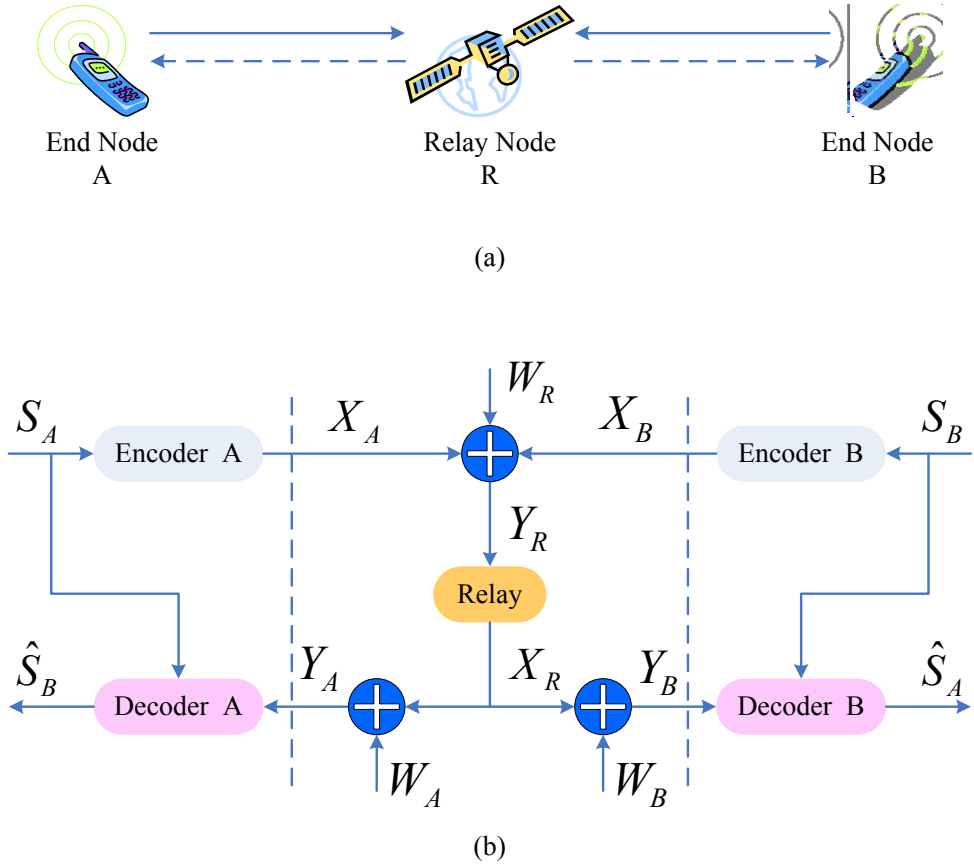}
\caption{System model for two way relay channel.} \label{fig:system}
\end{figure}

We study the two-way relay channel as shown in Fig. \ref{fig:system}, in which nodes $A$ and $B$ exchange information with the help of relay node $R$. We assume that all nodes are half-duplex, i.e., a node cannot receive and transmit simultaneously. We also assume that there is no direct link between nodes $A$ and $B$. An example in practice is a satellite communication system in which the two end nodes on the earth can only communicate with each other via the relay satellite, as shown in Fig. \ref{fig:system}(a).

We consider a two-phase transmission scheme consisting of an uplink phase and a downlink phase. In the uplink phase, nodes $A$ and
$B$ transmit packets to node $R$ simultaneously. In the downlink phase, based on the overlapped signals received from $A$ and $B$, $R$ constructs a network-coded packet and broadcast the packet to $A$ and $B$. Upon receiving the network-coded packet, $A$ ($B$) then attempts to recover the original packet transmitted by $B$ ($A$) in the uplink phase using self-information \cite{PNC06}.

This paper focuses on the performance of the uplink phase because the performance of the downlink phase is similar to that in a conventional point-to-point link. Consider the uplink phase. If only $A$ transmitted, then the received complex baseband signal at $R$ (i.e., the received signal after down-conversion from the carrier frequency and low-pass filtering) would be
\begin{align}
y_R (t) = \sum\limits_{n = 1}^N {h_A x_A [n]p(t - nT)}  + w_R (t),
\label{equ:ApncSys1}
\end{align}
where $h_A  = \sqrt {P_A } $ is the received signal amplitude; $(x_A [n])_{n = 1,...,N} $ are the symbols in the packet of $A$; $p(t
- nT)$ is the pulse shaping function for the baseband signal; and $w_R (t)$ is additive white Gaussian noise (AWGN) with double-sided power spectral density $S_w (f) = {{N_0 } \mathord{\left/{\vphantom {{N_0 } 2}} \right. \kern-\nulldelimiterspace} 2}$ for both the real and imaginary components within the baseband of interest. In the PNC set-up, $A$ and $B$ transmit simultaneously. In this case, the received complex baseband signal at $R$ is
\begin{align}
y_R (t) = \sum\limits_{n = 1}^N {\left\{ {h_A x_A [n]p(t - nT) + h_B
x_B [n]p(t - \Delta  - nT)} \right\}} + w_R (t),
\label{equ:ApncSys2}
\end{align}
where $h_B  = \sqrt {P_B } e^{j\phi }$ ($\phi$ is the relative phase offset between the signals from $A$ and $B$ due to phase asynchrony in their carrier-frequency oscillators and the difference in the path delays of the two uplink channels); $(x_B [n])_{n = 1,...,N}$ are
the symbols in the packet of $B$; and $\Delta$ is a time offset between the arrivals of the signals from $A$ and $B$. Without loss of generality, we assume the signal of $A$ arrives earlier than $B$. Furthermore, we assume $\Delta$ is within one symbol period \emph{T}. Thus, $0 \le \Delta  < T$.\footnote{If $\Delta$ is more than one symbol period, we could generalize our treatment here so that $N$ is larger than the number of symbols in a packet. The packets will only be partially overlapping, with non-overlapping symbols at the front end and tail end. Essentially, our assumption of $\Delta$ being within one symbol period implies that we are looking at the ``worst case'' with maximum overlapping between the two packets. When there are additional non-overlapping symbols at the front and tail ends, the decoding will have better error probability performance.} We refer to $\Delta$ and $\phi$ as the symbol and phase offsets (or misalignments) at $R$, respectively. For simplicity, we assume power control (equalization) so that $P_A  = P_B  = P$. Furthermore, for convenience, we assume time is expressed in unit of symbol duration, so that $T = 1$. Then, we can rewrite (\ref{equ:ApncSys2}) as
\begin{align}
y_R (t) = \sqrt P \sum\limits_{n = 1}^N {\left\{ {x_A [n]p(t - n) + x_B [n]p(t - \Delta  - n)e^{j\phi } } \right\}} + w_R (t).
\label{equ:ApncSys3}
\end{align}

In general, the pulse shaping function $p(t)$ can take different forms. The discussion on different pulse shaping functions, however, is beyond the scope of this paper. To bring out the essence of our results in the simplest manner, throughout this paper, we assume the rectangular pulse shape: $p(t) = rect(t) = u(t + 1) - u(t)$.

A critical design issue is how relay $R$ makes use of $y_R (t)$ to construct a packet for broadcast to nodes $A$ and $B$ in the downlink phase. In this paper, we assume that $R$ first oversamples $y_R (t)$ to obtain $2N + 1$ signal samples. It then uses the $2N + 1$ signal samples to construct an \emph{N}-symbol network-coded packet for broadcast to $A$ and $B$. The matched filtering and oversampling procedure is similar to that in \cite{CCRESM,UPNCICC11} and described below.
For $n=1,...,N,$
\begin{small} 
\begin{align}
y_R [2n - 1] &= \frac{1}{\Delta}\int_{(n - 1)}^{(n - 1) + \Delta } {\left( {x_A [n] + x_B [n - 1]e^{j\phi }  + \frac{w_R (t)}{\sqrt{P}}} \right)d} t = x_A [n] + x_B [n - 1]e^{j\phi }  + w_R [2n - 1], \nonumber\\
y_R [2n] &= \frac{1}{{1 - \Delta } }\int_{(n - 1) + \Delta }^n {\left( {x_A [n] + x_B [n]e^{j\phi }  + \frac{w_R (t)}{\sqrt{P}}} \right)d} t
= x_A [n] + x_B [n]e^{j\phi }  + w_R [2n], \label{equ:ApncSys4} \\
\text{and}\ \ \ \ \ \ \ \ \ \ &\ \nonumber\\
y_R [2N + 1] &= \frac{1}{\Delta }\int_N^{N + \Delta } {\left( {x_B [N]e^{j\phi }  + \frac{w_R (t)}{\sqrt{P}}} \right)d} t
= x_B [N]e^{j\phi }  + w_R [2N + 1], \nonumber
\end{align}
\end{small}
where $x_B [0] = 0$, and $w_R [2n-1]$ (also $w_R [2N+1]$) and $w_R [2n]$ are a zero-mean complex Gaussian noise with variance $N_0 /(2P\Delta )$ and $N_0 /\left( {2P(1 - \Delta )} \right)$, respectively, for both the real and imaginary components. Note that the powers in $x_A[n]$ and $x_B[n]$ have been respectively normalized to one unit, and $P\Delta/N_0 = E_s/N_0$ is SNR per symbol from node $A$ or node $B$. Based on $\left(y_R[n]\right)_{n=1,\dots,2N+1}$, relay $R$ constructs a network-coded packet $\left(x_R[n]\right)_{n=1,\dots,N}$ for broadcast to end nodes $A$ and $B$

Fig. \ref{fig:system}(b) is a schematic diagram. It incorporates both channel coding and physical-layer network coding. This paper adopts the following notation:
\begin{itemize}
\item $S_i  = \left( {s_i [1],s_i [2],\dots,s_i [M]} \right)$ denotes the source packet of node \emph{i}, $i \in \{ A,B\}$;

\item $X_i  = \left( {x_i [1],x_i [2],\dots,x_i [N]} \right)$ denotes the source packet of node \emph{i}, $i \in \{ A,B\}$;

\item $Y_R  = \left( {y_R [1],y_R [2],\dots,y_R [N],y_R [N + 1],...,y_R [2N + 1]} \right)$ denotes the received packet (with the aforementioned oversampling) at relay node $R$;

\item $W_R  = \left( {w_R [1],w_R [2],\dots,w_R [N],w_R [N + 1],...,\\w_R [2N + 1]} \right)$ denotes the receiver noise at node $R$;

\item $X_R  = \left( {x_R [1],x_R [2],\dots,x_R [N]} \right)$ denotes the network-coded packet at relay node $R$;

\item $Y_i  = \left( {y_i [1],y_i [2],\dots,y_i [N]} \right)$ denotes the received PNC packet at node \emph{i}, $i \in \{ A,B\}$;

\item $W_i  = \left( {w_i [1],w_i [2],\dots,w_i [N]} \right)$ denotes the receiver noise at node \emph{i}, $i \in \{ A,B\}$;


\item $\hat S_i  = \left( {\hat s_i [1],\hat s_i [2],\dots,\hat s_i [N]} \right)$ denotes the decoded source packet of node \emph{i}, $i \in \{A,B\}$, at the other end node;
\end{itemize}
where $M$ is the number of source symbols, and $N$ is the number of channel-coded symbols. Note here that, throughout this paper, we focus on BPSK and QPSK in our analytical and simulation results, although the framework can be extended to more complex constellations. For BPSK, $s_i [ \cdot ], x_i [ \cdot ],\hat s_i [ \cdot ],x_R [ \cdot ] \in \{-1,1\}$; for QPSK, $s_i [ \cdot ], x_i [ \cdot ],\hat s_i [ \cdot ],x_R [ \cdot ] \in \{ {{(1 + j)} \mathord{\left/ {\vphantom {{(1 + j)} {\sqrt 2 }}} \right. \kern-\nulldelimiterspace} {\sqrt 2 }},{{{\rm{ }}( - 1 + j)} \mathord{\left/ {\vphantom {{{\rm{ }}( - 1 + j)} {\sqrt 2 ,}}} \right. \kern-\nulldelimiterspace} {\sqrt 2 ,}}{\rm{ }}{{( - 1 - j)} \mathord{\left/ {\vphantom {{( - 1 - j)} {\sqrt 2 ,}}} \right. \kern-\nulldelimiterspace} {\sqrt 2 ,}}{\rm{ }}{{(1 - j)} \mathord{\left/ {\vphantom {{(1 - j)} {\sqrt 2 }}} \right. \kern-\nulldelimiterspace} {\sqrt 2 }}\}$, and $y_i [ \cdot ],y_R [ \cdot ],w_i [ \cdot ],w_R [ \cdot ] \in \mathbb{C}$.

\section{Unchannel-coded PNC}\label{Sec:UPNC}

This section focuses on unchannel-coded PNC, where each end node transmits the source information without channel coding. Thus, with respect to Fig. \ref{fig:system}(b), we have $X_A = S_A$ and $X_B = S_B$. For asynchronous unchannel-coded PNC, we investigate the use of Belief Propagation (BP) in the PNC mapping process to deal with phase and symbol asynchronies. We find that symbol misalignment can drastically reduce the performance penalty due to phase offset.

\subsection{Synchronous Unchannel-coded PNC} \label{Sec:UPNC1}
We first give a quick review of synchronous unchannel-coded PNC.

\begin{definition}[Synchronous Unchannel-coded PNC]
In synchronous unchannel-coded PNC, the two end nodes transmit their packets $X_A = S_A$ and $X_B = S_B$ (without channel coding) in a synchronous manner so that the relay node $R$ receives the combined signals with $\phi  = 0$ and $\Delta  = 0$. The received baseband packet at $R$ is $Y_R  = X_A  + X_B  + W_R$ with \emph{N} symbols. Node $R$ transforms $Y_R$ into a network-coded packet $X_R  = f(Y_R )$ for transmission in the downlink phase.
\end{definition}

For this case, with reference to (\ref{equ:ApncSys4}), since $\Delta  = 0$, the variance of the noise term $w_R [2n - 1]$ is infinite, and the signal is contained only in the even terms $y_R [2n]$. Thus, we can write
\begin{align}
y_R [2n] = x_A [n] + x_B [n] + w_R [2n], \label{equ:ApncBP1}
\end{align}
where $n = 1,...,N$, and $w_R [2n]$ is zero-mean Gaussian noise with variance $\sigma ^2  = N_0/(2P)$ for both the real and imaginary components.

For BPSK, $x_i [n] \in \{-1, 1\}$. Only the real component of $w_R [2n]$ needs to be considered. For QPSK, since we are considering a synchronous system, the in-phase and quadrature-phase components in (\ref{equ:ApncBP1}) are independent; it can therefore be considered as two parallel BPSK systems. Thus, in the following, we only consider BPSK.

Let us consider a particular time index \emph{n}, and omit the index \emph{n} in our notation for simplicity. The \emph{a posteriori} probability of the combination of source symbols $(x_A ,x_B)$ is given by
\begin{align}
&\Pr (x_A ,x_B |y_R ) = \frac{{\Pr (y_R |x_A ,x_B )}}{{4\Pr (y_R )}}
= \frac{P}{{4\Pr (y_R )\sqrt {2\pi \sigma ^2 } }}\exp \left\{ {\frac{{[y_R  - x_A  - x_B )]^2 }}{{2\sigma ^2/P }}} \right\}.\label{equ:ApncSync}
\end{align}

Let us use $x_i  = 1$ represent bit 0 and $x_i  = -1$ represent bit 1. Suppose that the downlink transmission also uses BPSK. For PNC output $x_R $, we assume we want to get the XOR mapping, i.e., $x_R  = x_A  \oplus x_B $ \cite{PNC06}. Then, $x_R  = 1$ if $x_A  = x_B $, and $x_R  =  - 1$ if $x_A  \ne x_B $. The following decision rule can be used to map $y_R$ to $x_R$:
\begin{align}
&\Pr (x_A  = 1,x_B  = 1|y_R ) + \Pr (x_A  =  - 1,x_B  =  - 1|y_R )\nonumber\\
&\mathop \gtrless \limits^{x_R  = 1} \limits_{x_R  = -1}
\Pr (x_A  = 1,x_B  =  - 1|y_R ) + \Pr (x_A  =  - 1,x_B  = 1|y_R ) \nonumber\\
\Rightarrow
&\left( {\exp \left\{ {\frac{{(y_R  - 2)^2 }}{{2\sigma ^2/P }}} \right\} + \exp \left\{ {\frac{{(y_R  + 2)^2 }}{{2\sigma ^2/P }}} \right\}} \right) \mathop \gtrless \limits^{x_R  = 1} \limits_{x_R  = -1} 2 \exp \left\{ {\frac{{y_R ^2 }}{{2\sigma ^2/P }}} \right\}. \label{equ:ApncBP2}
\end{align}

\subsection{BP-UPNC: A Belief Propagation based Unchannel-coded PNC Scheme} \label{Sec:UPNC2}
\begin{definition}[Asynchronous unchannel-coded PNC]
In asynchronous unchannel-coded PNC, the two end nodes transmit their packets $X_A = S_A$ and $X_B = S_B$ (without channel coding) in an asynchronous manner so that the relay node $R$ receives the combined signals with $\phi \ne 0$ and/or $\Delta \ne 0$. The received baseband packet at $R$ is $Y_R  = X_A  + X_B  + W_R$ with $2N+1$ symbols. Node $R$ transforms $Y_R$ into a network-coded packet $X_R  = f(Y_R)$ for transmission in the downlink phase.
\end{definition}

This subsection presents an unchannel-coded PNC decoding scheme for the mapping $X_R = f(Y_R)$, based on belief propagation that deals with symbol and phase asynchronies jointly. We refer to the method as BP-UPNC. We make use of the oversampled symbols in (\ref{equ:ApncSys4}) to construct a Tanner graph \cite{KschIT01} as shown in Fig. \ref{fig:TannerBPPNC}. In the Tanner graph, $Y_R$ denotes the evidence nodes, and there are $2N+1$ such nodes; $\Psi$ denotes the constraint nodes (also known as the compatibility or check nodes); and \emph{X} denotes the source nodes (also known as the variable nodes). For simplicity, we use $x^{i,j}$ to denote $\left( {x_A [i],x_B [j]} \right)$. The correlation between two adjacent joint symbols is modeled by the compatibility functions (i.e., check nodes) $\psi _o (x^{n,n - 1} ,x^{n,n} )$ and $\psi _e (x^{n,n} ,x^{n + 1,n} )$ for the odd and even compatibility nodes:
\begin{equation}
\begin{array}{l}
\psi_o (x^{n,n - 1} ,x^{n,n} ) = \left\{ \begin{array}{ll}
 1 \ \ \text{if}\ x_A [n]\ \text{in}\ x^{n,n - 1}\ \text{and}\ x^{n,n}\ \text{are equal}\\
 0 \ \ \text{otherwise}\\
 \end{array} \right.\\
\psi_e (x^{n,n} ,x^{n+1,n} ) = \left\{ \begin{array}{ll}
 1 \ \ \text{if}\ x_B [n]\ \text{in}\ x^{n,n}\ \text{and}\ x^{n+1,n}\ \text{are equal}\\
 0 \ \ \text{otherwise}\\
 \end{array} \right.
\end{array}
\end{equation}
We first decode the combination $\left( {x_A [n],x_B [n]} \right)$ in $X$. Note that the Tanner graph has a tree structure. This means that BP can find the ``\emph{exact}'' \emph{a posteriori} probability $P(x^{n,n} |Y_R )$ for $n=1,\dots,N$. Furthermore, the solution can be found after only one iteration of the message-passing algorithm \cite{KschIT01}. From the decoded $P(x^{n,n} |Y_R)$, we can then find the maximum \emph{a posteriori} probability (MAP) XOR value
\begin{align}
x_R [n] =\arg \mathop {\max }\limits_{x}  {P \left( {x_A[n] \oplus x_B[n] = x|Y_R } \right)}
= \arg \mathop {\max }\limits_{x} \sum\limits_{\scriptstyle x^{n,n} :\ \scriptstyle x_A [n] \oplus x_B [n] = x} {P \left( {x^{n,n} |Y_R } \right)}. \label{equ:UpncDec}
\end{align}

In summary, BP can converge quickly and is MAP-optimal as far as the BER of $x_A [n] \oplus x_B [n]$ is concerned. Note that MAP optimal is also ML optimal here because the \emph{a priori} probability $P(x^{n,n})$ for different values of $x^{n,n}$ are equally likely.


\vspace{0.08in}
\subsubsection{BP-UPNC Design}
Let us consider QPSK modulation, in particular, we define $\chi = \left\{ 1 + j, - 1 + j,\right.$ $\left. - 1 - j,1 - j \right\}$ as the symbol set. With reference to (\ref{equ:ApncSys4}), we have $x_A [n] = {a \mathord{\left/ {\vphantom {a {\sqrt 2 }}} \right. \kern-\nulldelimiterspace} {\sqrt 2 }}$ and $x_B [n] = {b \mathord{\left/ {\vphantom {b {\sqrt 2 }}} \right. } {\sqrt 2}}$, where $a,b \in \chi$. Define $p_k^{a,b}  = P\left( x_A [\left\lceil {k/2} \right\rceil ] = {a \mathord{\left/ {\vphantom {a {\sqrt 2 }}} \right. } {\sqrt 2 }},x_B [\left\lfloor {k/2} \right\rfloor ]\right. \\\left.= {b \mathord{\left/ {\vphantom {b {\sqrt 2 }}} \right. \kern-\nulldelimiterspace} {\sqrt 2 }}\ \big|y_R [k] \right)$. Note that here, $p_k^{a,b}$ is computed based on $y_R [k]$ only, and not on the whole $Y_R $. Also, $p_k^{a,b}$ is fixed and does not change throughout the message passing algorithm in the Tanner graph. $P_{2n - 1}^{a,b}$ and $P_{2n}^{a,b}$, $n = 1,2,\dots,N$, are given as follows:
\begin{small} 
\begin{align}
&p_{2n - 1}^{a,b}  = P\left(x_A [n] = \frac{a}{{\sqrt 2 }},x_B [n - 1] = \frac{b}{{\sqrt 2 }} \bigg\vert y_R [2n - 1] \right) \nonumber\\
&= \frac{P}{{2\pi \sigma ^2 /\Delta }}\exp \left\{ {\frac{{\left( {y_R^{{\mathop{\rm Re}\nolimits} } [2n - 1] - {{{\mathop{\rm Re}\nolimits} \left( {a + be^{j\phi } } \right)} \mathord{\left/
{\vphantom {{{\mathop{\rm Re}\nolimits} \left( {a + be^{j\phi } } \right)} {\sqrt 2 }}} \right.
\kern-\nulldelimiterspace} {\sqrt 2 }}} \right)^2 }}{{2\sigma ^2 /(\Delta P) }}} \right\}\cdot \exp \left\{ {\frac{{\left( {y_R^{{\mathop{\rm Im}\nolimits} } [2n - 1] - {{{\mathop{\rm Im}\nolimits} \left( {a + be^{j\phi } } \right)} \mathord{\left/
{\vphantom {{{\mathop{\rm Im}\nolimits} \left( {a + be^{j\phi } } \right)} {\sqrt 2 }}} \right.
\kern-\nulldelimiterspace} {\sqrt 2 }}} \right)^2 }}{{2\sigma ^2 /(\Delta P) }}} \right\}, \label{equ:ApncBPIni1}
\end{align}
\begin{align}
&p_{2n}^{a,b}  = P\left(x_A [n] = \frac{a}{{\sqrt 2 }},x_B [n] = \frac{b}{{\sqrt 2 }} \bigg\vert y_R [2n] \right) \nonumber\\
&= \frac{P}{{2\pi \sigma ^2 /(1 - \Delta )}}
\exp \left\{ {\frac{{\left( {y_R^{{\mathop{\rm Re}\nolimits} } [2n] - {{{\mathop{\rm Re}\nolimits} \left( {a + be^{j\phi } } \right)} \mathord{\left/
{\vphantom {{{\mathop{\rm Re}\nolimits} \left( {a + be^{j\phi } } \right)} {\sqrt 2 }}} \right.
\kern-\nulldelimiterspace} {\sqrt 2 }}} \right)^2 }}{{2\sigma ^2 /[(1 - \Delta )P] }}} \right\}\cdot \exp \left\{ {\frac{{\left( {y_R^{{\mathop{\rm Im}\nolimits} } [2n] - {{{\mathop{\rm Im}\nolimits} \left( {a + be^{j\phi } } \right)} \mathord{\left/
{\vphantom {{{\mathop{\rm Im}\nolimits} \left( {a + be^{j\phi } } \right)} {\sqrt 2 }}} \right.
\kern-\nulldelimiterspace} {\sqrt 2 }}} \right)^2 }}{{2\sigma ^2 /[(1 - \Delta )P] }}} \right\}. \label{equ:ApncBPIni2}
\end{align}
\end{small} 
Note that except for the first and last symbols, each of $p_{2n - 1}^{a,b}$ and $p_{2n}^{a,b}$ has 16 possible combinations (4 possibilities for $a$ and 4 possibilities for $b$). The first and last symbols have 4 possibilities, as follows:
$ p_1^{a,0}  = P(x_A [N + 1] = {a \mathord{\left/
{\vphantom {a {\sqrt 2 }}} \right.
\kern-\nulldelimiterspace} {\sqrt 2 }},x_B [n] = 0|y_R [1])
= \frac{P}{{2\pi \sigma ^2 /\Delta }}\exp \left\{ {\frac{{\left( {y_R^{{\mathop{\rm Re}\nolimits} } [1] - {\mathop{\rm Re}\nolimits} \left( {{a \mathord{\left/
{\vphantom {a {\sqrt 2 }}} \right.
\kern-\nulldelimiterspace} {\sqrt 2 }}} \right)} \right)^2 }}{{2\sigma ^2 /(\Delta P) }}} \right\} \cdot
\exp \left\{ {\frac{{\left( {y_R^{{\mathop{\rm Im}\nolimits} } [1] - {\mathop{\rm Im}\nolimits} \left( {{a \mathord{\left/
{\vphantom {a {\sqrt 2 }}} \right.
\kern-\nulldelimiterspace} {\sqrt 2 }}} \right)} \right)^2 }}{{2\sigma ^2 /(\Delta P) }}} \right\},$
and
\begin{small}
$ p_{2N + 1}^{0,b}  = P(x_A [N + 1] = 0,x_B [n] = {b \mathord{\left/
{\vphantom {b {\sqrt 2 }}} \right.
\kern-\nulldelimiterspace} {\sqrt 2 }}|y_R [2N + 1])
= \frac{P}{{2\pi \sigma ^2 /\Delta }} \exp \left\{ {\frac{{\left( {y_R^{{\mathop{\rm Re}\nolimits} } [2N + 1] - {\mathop{\rm Re}\nolimits} \left( {{{be^{j\phi } } \mathord{\left/
{\vphantom {{be^{j\phi } } {\sqrt 2 }}} \right.
\kern-\nulldelimiterspace} {\sqrt 2 }}} \right)} \right)^2 }}{{2\sigma ^2 /(\Delta P) }}} \right\}
\exp \left\{ {\frac{{\left( {y_R^{{\mathop{\rm Im}\nolimits} } [2N + 1] - {\mathop{\rm Im}\nolimits} \left( {{{be^{j\phi } } \mathord{\left/
{\vphantom {{be^{j\phi } } {\sqrt 2 }}} \right.
\kern-\nulldelimiterspace} {\sqrt 2 }}} \right)} \right)^2 }}{{2\sigma ^2 /(\Delta P) }}} \right\}.$
\end{small}
\subsubsection{Message Update Rules}

\begin{figure}[h]
\centering
\includegraphics[width=1\textwidth]{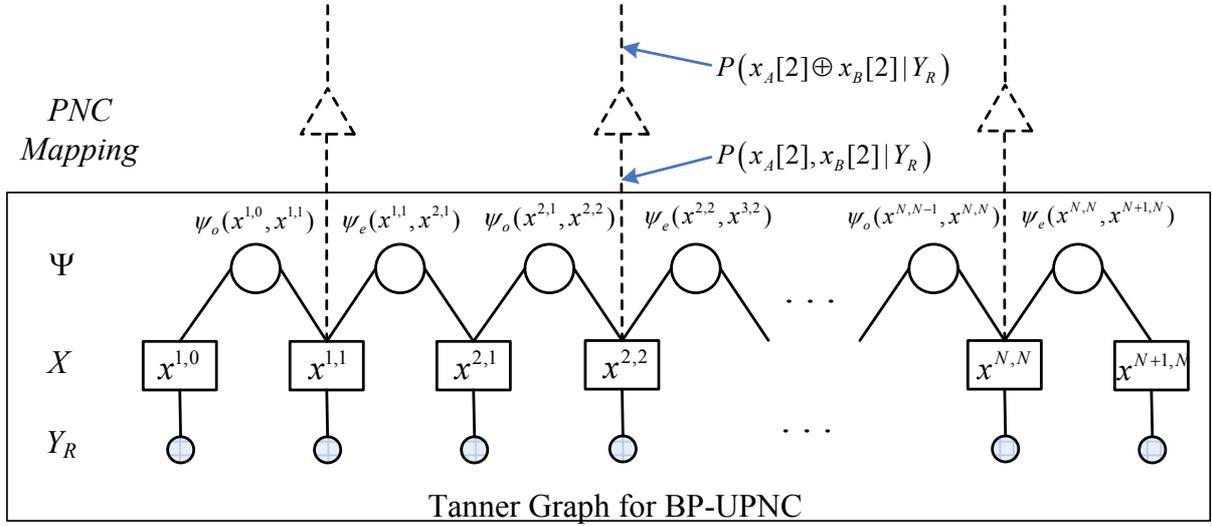}
\caption{Tanner graph of the BP-UPNC and PNC mapping after BP algorithm on the Tanner graph. The triangle nodes perform the PNC mapping described in Section \ref{Sec:UPNC2}.} \label{fig:TannerBPPNC}
\end{figure}

Given the evidence node values computed by (\ref{equ:ApncBPIni1}) and (\ref{equ:ApncBPIni2}), we now derive the message update rules for BP-UPNC. Since the Tanner graph (in Fig. \ref{fig:TannerBPPNC}) for BP-UPNC has a tree structure, the decoding of the joint probability   can be done by passing the messages only once on each direction of an edge \cite{YedTR01}. As described below, we could consider the right-bound messages first followed by the left-bound messages.

\begin{figure}[h]
\centering
\includegraphics[width=1\textwidth]{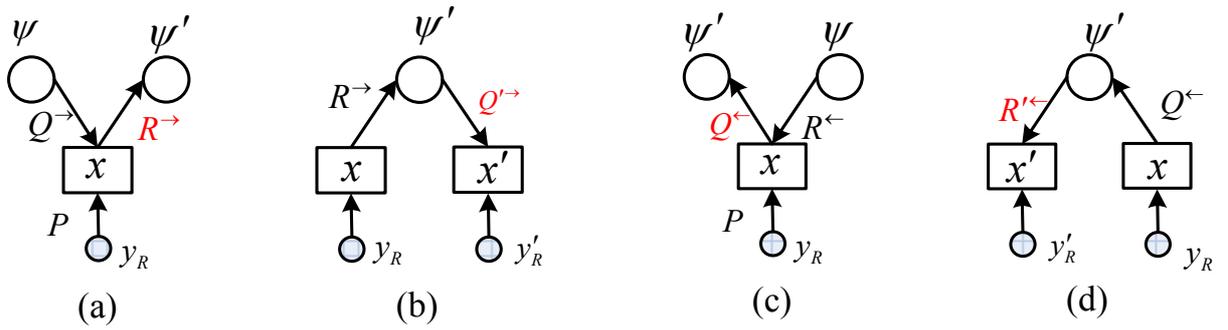}
\caption{BP-UPNC message update procedure. (a) and (b): Update of messages from left to right; (c) and (d): Update of messages from right to left.} \label{fig:MessagePass}
\end{figure}

We represent messages on the edges with respect to the compatibility nodes $\Psi$ in Fig. \ref{fig:TannerBPPNC} by $Q_k$ and $R_k$. Specifically, $Q_k$ ($R_k$) is the message connected to the right (left) of the $k$-th compatibility node, as illustrated in Fig. \ref{fig:MessagePass}. $P_k  = (p_k^{1 + j,1 + j} ,p_k^{1 + j, - 1 + j} ,\dots,p_k^{1 - j,1 - j} )$ is a $16 \times 1$ probability vector associated with the $k$-th evidence node, $y_R[k]$, where each component $p_k^{a,b}  = P\left( x_A [\left\lceil {k/2} \right\rceil ] = {a \mathord{\left/ {\vphantom {a {\sqrt 2 }}} \right. } {\sqrt 2 }},x_B [\left\lfloor {k/2} \right\rfloor ]\right. \left.= {b \mathord{\left/ {\vphantom {b {\sqrt 2 }}} \right. \kern-\nulldelimiterspace} {\sqrt 2 }}\ \big|y_R [k] \right)$ is the joint conditional probability of $\left(x_A [\left\lceil {k/2} \right\rceil ],x_B [\left\lfloor {k/2} \right\rfloor ]\right)$ given $y_R [k]$ of a particular $(a,b)$ combination.


We omit the index $k$ to avoid cluttering in the following discussion of message-update rules. We follow the principles and assumptions of the BP algorithm to derive the update equations, i.e., the output of a node should be consistent with the inputs while adopting a ``sum of product'' format of the possible input combinations \cite{YedTR01}.

\noindent\emph{Step 1. Update of right-bound messages:}

With reference to Fig. \ref{fig:MessagePass}(a), we derive the update equations for the right-bound message $R^ \to   = (r^{1 + j,1 + j} ,r^{1 + j, - 1 + j} ,\dots,r^{1 - j,1 - j} )$ based on the right-bound message  and the fixed message $Q^ \to   = (q^{1 + j,1 + j} ,q^{1 + j, - 1 + j} ,\dots,q^{1 - j,1 - j} )$ from the evidence node $P = (p^{1 + j,1 + j} ,p^{1 + j, - 1 + j}\\ ,\dots,p^{1 - j,1 - j} )$. Based on the sum-product principle of the BP algorithm, we have
\begin{align}
r^{a,b}  = p^{a,b} q^{a,b}. \label{equ:ApncBPUpdate1}
\end{align}
For the input message into the leftmost compatibility node, however, we should modify (\ref{equ:ApncBPUpdate1}) with $r^{a,b}  = p^{a,b}$.

With reference to Fig. \ref{fig:MessagePass}(b) we derive the update equations for the message ${Q}'^\to   = (q'^{1 + j,1 + j} ,q'^{1 + j, - 1 + j} ,\dots,q'^{1 - j,1 - j} )$ based on the input message $R^{\to} = (r^{1 + j,1 + j}, r^{1 + j, - 1 + j}, \dots,  \\  r^{1 - j,1 - j} )$ of the compatibility node. Note that for $x = x^{n,n}$ and $x' = x^{n + 1,n}$, the common symbol overlapping in two adjacent samples is $x_B [n]$, thus we have
\begin{align}
q'^{1 + j,b}  = q'^{ - 1 + j,b}  = q'^{ - 1 - j,b}  = q'^{1 - j,b} = \beta \sum\limits_{a \in \chi } {r^{a,b} }, \label{equ:ApncBPUpdate2}
\end{align}
where $b \in \chi $ and $\beta$ is a normalization factor to ensure that the sum of probabilities $\sum\limits_{a,b \in \chi} {q'^{a,b} }  = 1$.
For $x = x^{n + 1,n}$ and $x' = x^{n + 1,n + 1}$, the common symbol is $x_A [n + 1]$. Then, we have
\begin{align}
q'^{a,1 + j}  = q'^{a, - 1 + j}  = q'^{a, - 1 - j}  = q'^{a,1 - j} = \beta \sum\limits_{b \in \chi } {r^{a,b} }, \label{equ:ApncBPUpdate3}
\end{align}
where $a \in \chi$ and $\beta$ is a normalization factor to ensure that the sum of probabilities $\sum\limits_{a,b \in \chi} {q'^{a,b} }  = 1$.


We repeat the procedures in Fig. \ref{fig:MessagePass}(a) and Fig. \ref{fig:MessagePass}(b) to compute all the successive right-bound messages, starting from the left to the right, until we reach the rightmost node.

\vspace*{.06in}
\noindent\emph{Step 2. Update of left-bound messages}

With reference to Fig. \ref{fig:MessagePass}(c) and (d), we use a similar procedure as in \emph{Step 1} to update the left-bound messages.

\vspace*{.06in}
After \emph{Step 1} and \emph{Step 2} above, the values of the messages converge, thanks to the tree structure of the Tanner graph. We then compute the 4-tuple.
\begin{small}
\begin{align}
\left(\sum\limits_{(a,b):a \oplus b = 1 + j} {p_{2n}^{a,b} q_{2n}^{a,b} r_{2n}^{a,b} } , \sum\limits_{(a,b):a \oplus b =  - 1 + j} {p_{2n}^{a,b} q_{2n}^{a,b} r_{2n}^{a,b} }, \sum\limits_{(a,b):a \oplus b =  - 1 - j} {p_{2n}^{a,b} q_{2n}^{a,b} r_{2n}^{a,b} } , \sum\limits_{(a,b):a \oplus b = 1 - j} {p_{2n}^{a,b} q_{2n}^{a,b} r_{2n}^{a,b} } \right), \nonumber
\end{align}
\end{small}
where $a,b \in \chi$. The maximum-likelihood ${x_A [n] \oplus x_B [n]}$ is given by the $a \oplus b$ that yields the largest element among the four elements. Note that we only use $P\left( {x_A [n],x_B [n]} | Y_R \right)$, and not $P\left( {x_A [n + 1],x_B [n]} | Y_R \right)$, to obtain $P\left( {x_A [n] \oplus x_B [n]} | Y_R \right)$, because the information required to get $P\left( {x_A [n] \oplus x_B [n]} | Y_R \right)$ is fully captured in $P\left( {x_A [n],x_B [n]} | Y_R \right)$.

\subsection{Numerical Results}\label{Sec:UPNC3}
This section presents simulation results for BP-UPNC. We compare the performance of asynchronous unchannel-coded PNC with that of the perfectly synchronized case \cite{PNC06}..

\subsubsection{Summary of Results}
Our simulations yield the following findings:
\begin{itemize}
\item For BPSK, the 3 dB BER performance penalty due to phase or symbol asynchrony using the decoding methods in \cite{PNC06, SyncPNC06} is reduced to less than 0.5 dB with our method.
\item For QPSK, the BER performance penalty due to phase asynchrony can be as high as 6-7 dB when the symbols are aligned, as can be seen from Fig. \ref{fig:BPPNCsimulationQPSK}(a). However, with half symbol misalignment, our method can reduce the penalty to less than 1 dB.
\end{itemize}

The general conclusion is that misalignment makes the system more robust against phase asynchrony. If one could control the symbol timings (e.g., \referred{RahulSourceSync10}\cite{RahulSourceSync10} presented a method to control the timings of symbols from different sources), it would be advantageous to deliberately introduce a half symbol offset in unchannel-coded PNC.


\subsubsection{Detailed Description}
\begin{figure}[h]
\centering
\includegraphics[width=1\textwidth]{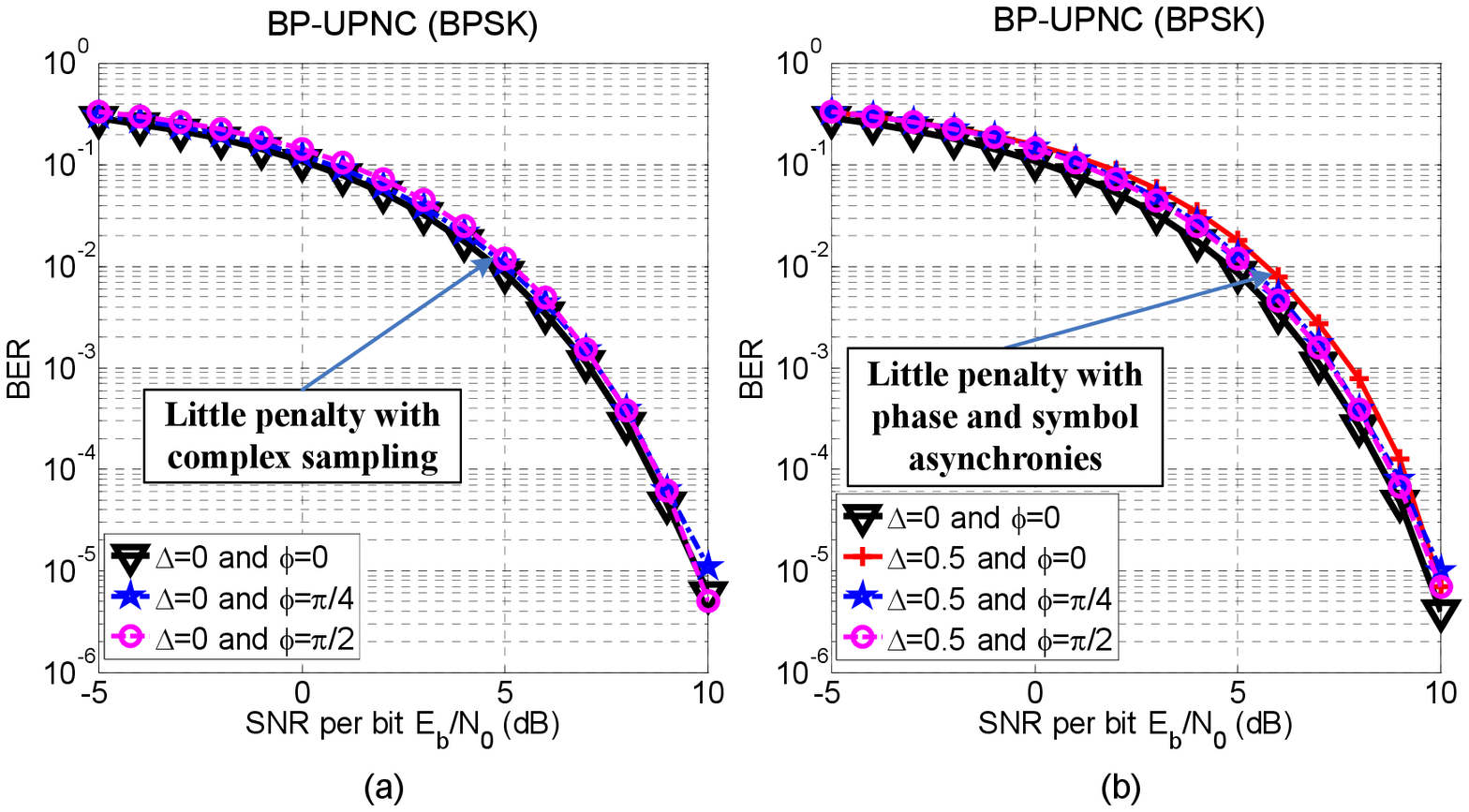}
\caption{BER of the uplink XORed value $x_A[n] \oplus x_B[n]$ in BP-UPNC for BPSK modulated unchannel-coded PNC. (a): BP-UPNC without symbol asynchrony ($\Delta = 0$); (b): BP-UPNC with symbol asynchrony ($\Delta \ne 0$). Note that $s_A[n] = x_A[n]$ and $s_B[n] = x_B[n]$ for unchannel-coded PNC.}
\label{fig:BPPNCsimulationBPSK}
\end{figure}

\begin{figure}[h]
\centering
\includegraphics[width=1\textwidth]{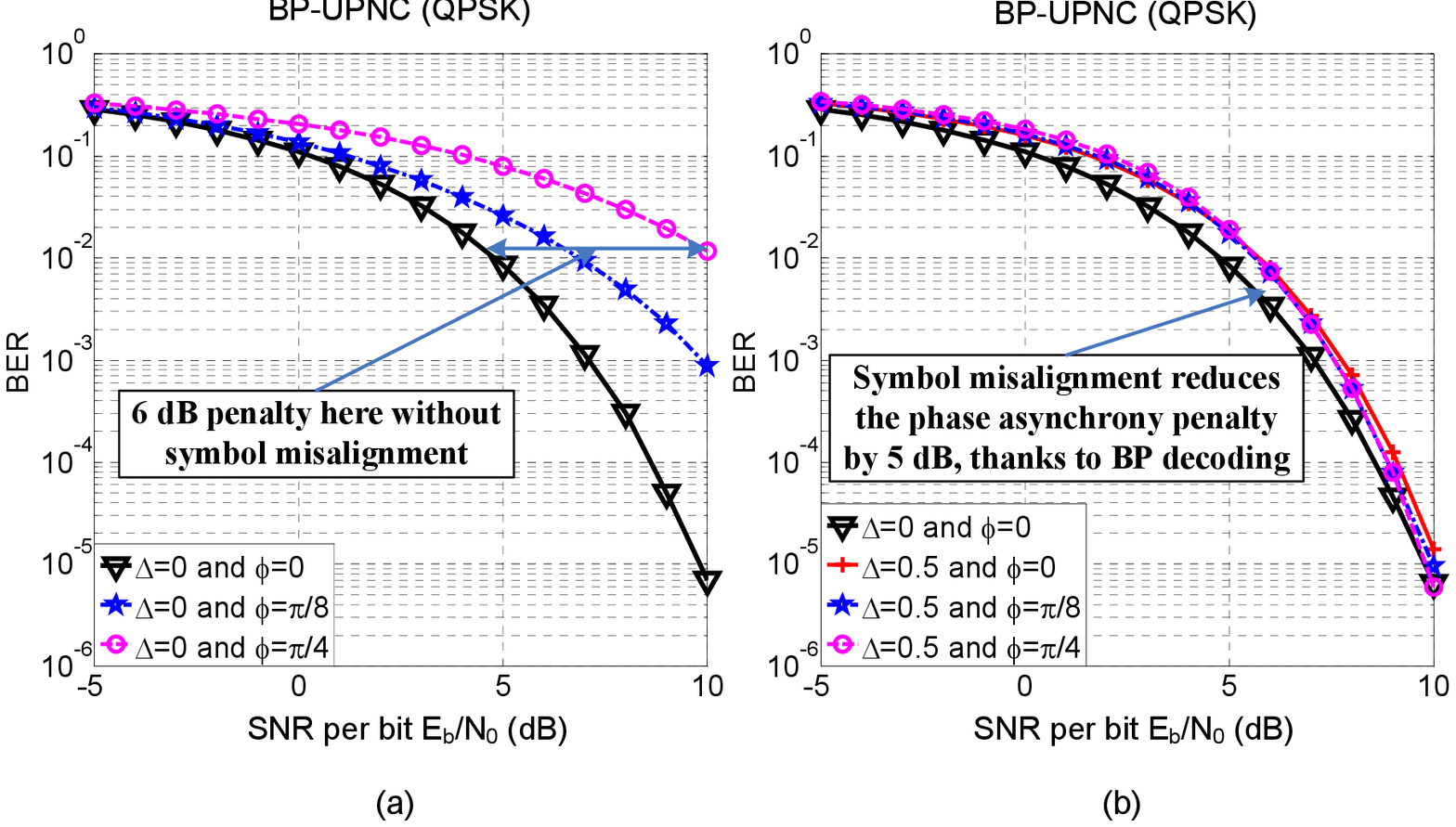}
\caption{BER of the uplink XORed value $x_A[n] \oplus x_B[n]$ in BP-UPNC for QPSK modulated unchannel-coded PNC. (a): BP-UPNC without symbol asynchrony ($\Delta = 0$); (b): BP-UPNC with symbol asynchrony ($\Delta \ne 0$). Note that $s_A[n] = x_A[n]$ and $s_B[n] = x_B[n]$ for unchannel-coded PNC.}
\label{fig:BPPNCsimulationQPSK}
\end{figure}

Fig. \ref{fig:BPPNCsimulationBPSK} and Fig. \ref{fig:BPPNCsimulationQPSK} show the simulation results for BPSK and QPSK, respectively. The \emph{x}-axis is the average SNR per bit of both end nodes, and the \emph{y}-axis is the BER for the uplink XORed value $S_A  \oplus S_B $.


For each data point, we simulate 10,000 packets of 2,048 bits. We use the synchronous unchannel-coded PNC as a benchmark to evaluate BP-UPNC. As can be seen from Fig. \ref{fig:BPPNCsimulationBPSK}(a) and (b), the BER performance penalties due to phase and symbol asynchronies are less than 0.5 dB for all SNR regimes for BPSK. That is, for BPSK, BP-UPNC reduces the BER performance penalty from 3 dB in \cite{PNC06} to only 0.5 dB. Note that BP-UPNC uses complex sampling that samples both the In-phase and Quadrature-phase components of the signal, whereas the scheme in \cite{PNC06} only samples in one dimension. For optimality, complex sampling is required in asynchronous PNC because of the phase offset.

For QPSK with symbol synchrony but phase asynchrony, as can be seen from Fig. \ref{fig:BPPNCsimulationQPSK}(a), the BER performance penalty can be as large as 6 to 7 dB. However, with half symbol misalignment, as can be seen from Fig. \ref{fig:BPPNCsimulationQPSK}(b), BP-UPNC reduces the penalty to within 1 dB (compared with the benchmark case where symbol and phase are perfectly synchronized).  In other words, symbol asynchrony can ameliorate the penalty due to phase asynchrony. This can be explained by the ``diversity and certainty propagation'' effects elaborated in the next subsection. In addition, we note from Fig. \ref{fig:BPPNCsimulationQPSK}(b) that when there is a half symbol offset, the phase offset effect becomes much less significant. Specifically, the spread of SNRs for a fixed BER under different phase offsets is less than 0.5 dB.
\vspace*{-0.11in}

\subsection{Diversity and Certainty Propagation}\label{CertProp}
\begin{figure}[h]
\centering
\includegraphics[width=0.95\textwidth]{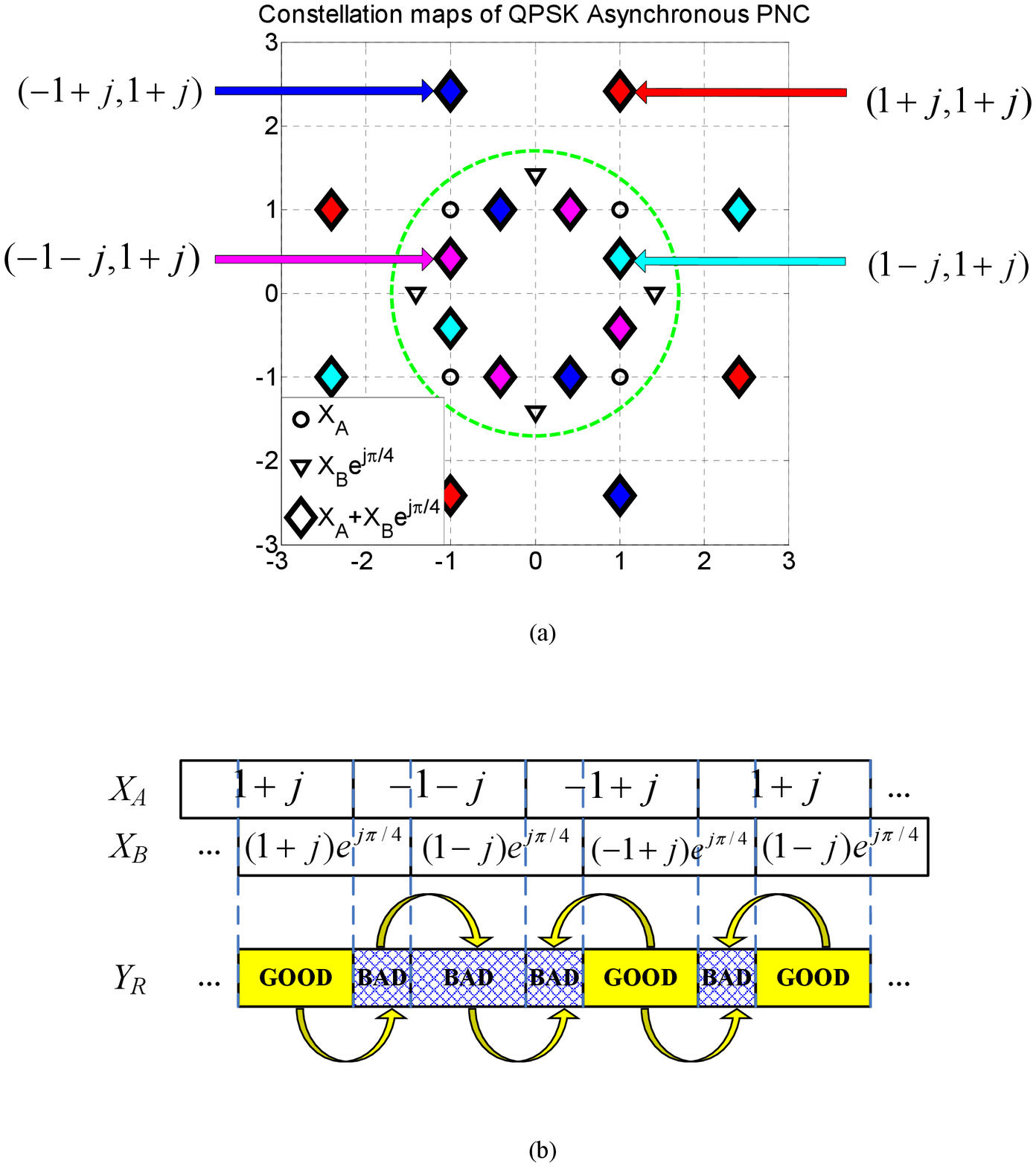}
\caption{Certainty propagation illustration. (a) Constellation map for phase asynchronous QPSK PNC (with symbol synchronization). There are four XORed PNC symbols for QPSK. Constellation points of the same color in the figure should be mapped to a same PNC symbol. And the amplitude of each symbol is $\sqrt 2$; (b) An illustration of certainty propagation. The yellow symbols denote the good constellation points and the white symbols with grids denote the bad constellation points of (a), respectively.} \label{fig:ConstMapAndCertProp}
\end{figure}

In QPSK, each symbol has four possible values. Thus, the joint symbol from both sources has 16 possible values. The constellation map of the joint symbol varies according to the phase offset. Fig. \ref{fig:ConstMapAndCertProp}(a) shows the constellation map of a joint symbol when the phase offset is $\pi/4$, where the 16 diamonds corresponding to the 16 possibilities. For example, a point with value $1 + (1 - \sqrt 2 )j$ corresponds to the joint symbol $x^{n,n}  = \left( {1 + j, - 1 - j} \right)$ in Fig. \ref{fig:ConstMapAndCertProp}(a) due to the  phase shift (i.e., $1 + (1 - \sqrt 2 )j = (1 + j) + ( - 1 - j)e^{j{\pi  \mathord{\left/ {\vphantom {\pi  4}} \right. \kern-\nulldelimiterspace} 4}}$). In PNC, the 16 possibilities need to be mapped to four XOR possibilities for the PNC symbol. In Fig. \ref{fig:ConstMapAndCertProp}(a), the diamonds are grouped into groups of four different colors. The diamonds of the same color are to be mapped to the same XOR PNC symbol according to $x_A  \oplus x_B$. In this mapping process, some of the constellation points are more prone to errors than other constellation points, and the BER is dominated by these bad constellation points.

With reference to Fig. \ref{fig:ConstMapAndCertProp}(a), the eight diamonds within the green circle are ``bad constellation points''. Adjacent points among the 8 points are mapped to different XOR values, but the distance between two adjacent points is small. By contrast, the eight points outside the green circle are ``good constellation points'' because the distance between adjacent points is large. When symbols are synchronized (i.e., $\Delta  = 0$), there are altogether $N$ joint symbols. On average, half of them will be bad constellation points with high BER.

Now, consider what if there is a symbol offset, say $\Delta  = 0.5$. We have $2N+1$ joint symbols, out of which about $N$ will be good constellation points. A symbol from a source is combined with two symbols from the other source in two joint symbols received at the relay. Both the joint symbols have to be bad for poor performance. Thus, the diversity itself may give some improvement. In addition, there is a certainty propagation effect, as explained below.

Consider a good constellation point, say $x^{n,n}  = \left( {x_A [n],x_B [n]} \right)$. For this point, we may be able to decode not just the XOR, but the individual values of $x_A [n]$ and $x_B [n]$ with high certainty. Now, suppose that the next joint symbol $x^{n{\rm{ + 1}},n}  = (x_A [n{\rm{ + 1}}],x_B [n])$ is a bad constellation point. But since we have good certainty about $x_B [n]$ from the previous good constellation point, the uncertainly in $x^{n + 1,n}$ can be reduced by the certainty propagated from $x^{n,n}$ with the BP algorithm. In other words, once $x_B [n]$ is known, $x^{n + 1,n}$ is not a bad constellation point anymore. Certainty can propagate along successive symbols from left to right, as well as from right to left, as shown in Fig. \ref{fig:ConstMapAndCertProp}(b), significantly reducing BER.

\section{Channel-coded PNC} \label{Sec:CPNC}
This section generalizes the BP algorithm for application in channel-coded PNC. In the last section, we show that the BP algorithm can exploit symbol offset in unchannel-coded PNC to reduce the phase asynchrony penalty. A natural question is whether such improvement carries over to channel-coded PNC. Interestingly, we find that given the appropriate decoding scheme, symbol offset in channel-coded PNC has an even larger positive effect, so much so that the phase penalty becomes a phase reward in that phase asynchrony actually improves performance. More importantly, we find that channel coding has the effect of making the system performance a lot less sensitive to the symbol and phase asynchronies.

The structure of channel-coded PNC is shown in Fig. \ref{fig:system}(b). The channel inputs $X_i$ are the channel-coded symbols constructed by performing channel coding operation $\Gamma_i$ on the source symbols $S_i$, $i \in \{A,B\}$. Each source node has $M$ symbols with a coding rate of $M/N$. The relay $R$ performs matched-filtering and sampling to get $Y_R$, which has $N$ symbols for Cases 1 and 3 in Table \ref{tab:Class}, and $2N + 1$ oversampled symbols for for Cases 2 and 4 in Table \ref{tab:Class}. Then, it transforms $Y_R$ to a channel-coded network-coded packet $X_R$ with $N$ symbols.

In this paper, we assume all nodes use the same channel code. That is $\Gamma_i = \Gamma$, $\forall i \in \{A, B, R\}$. In particular, we use the Repeat Accumulate (RA) channel code \referred{JinThesis01, ShengliJSAC09, C-CRESM}\cite{JinThesis01, ShengliJSAC09, CCRESM} to illustrate our main ideas and for our numerical studies. Other channel codes amenable to decoding by the BP algorithm can also be used.

For unchannel-coded PNC, thanks to the tree structure of the Tanner graph, we have an exact ML optimal BP algorithm, BP-UPNC. For channel-coded PNC, as will be seen, the channel code introduces loops in the Tanner graph. As a result, BP is only approximately ML optimal  \referred{YedTR01}\cite{YedTR01}. Furthermore, there are several ways to construct BP-CPNC (belief-propagation based channel-coded PNC algorithms). In this paper, we study two such methods, Jt-CNC and XOR-CD (to be detailed in Sections \ref{Sec:CPNC2} and \ref{Sec:CPNC3}).

\subsection{Channel-decoding and Network-Coding (CNC) Process}\label{Sec:CPNC1}
Recall that we wish to perform network coding on the received overlapped channel-coded packets. Specifically, based on the received signal $Y_R$, the relay wants to produce an output packet $X_R = f(Y_R)$ for broadcast to nodes $A$ and $B$. In particular, we would like $X_R$ to be the channel coded XOR of the source packets of nodes $A$ and $B$; i.e., $X_R$ is an estimate of $\Gamma(S_A \oplus S_B)$.

We could get $X_R$ directly from the received packet $Y_R$ without first channel decoding the source packets (i.e., we simply map $Y_R$ to $X_R$ on a symbol-by-symbol basis as in unchannel-coded PNC without performing channel decoding). We could also try to detect the XORed source packets $S_A \oplus S_B$ from $Y_R$, and then re-channel encode them to get $X_R$. In this paper, we consider the second method. The second method generally has better performance because the relay performs channel decoding to remove errors before forwarding the network-coded signal. The first method corresponds to end-to-end channel-coded PNC while the second method corresponds to link-by-link channel-coded PNC \referred{PNCSurvey,ShengliJSAC09}\cite{PNCSurvey,ShengliJSAC09}.

\begin{figure}[h]
\centering
\includegraphics[width=1\textwidth]{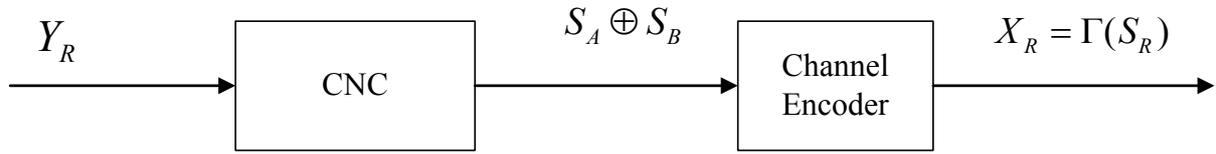}
\caption{Link-by-link channel-coded PNC, including the Channel-decoding and Network-Coding (CNC) process and the channel re-encoding process.}\label{fig:CNC}
\end{figure}

The basic idea in link-by-link channel-coded PNC is shown in Fig. \ref{fig:CNC}. It consists of two parts.


\emph{Part 1:}
The operation performed by the first part is referred to as the Channel-decoding and Network-Coding (CNC) process in \referred{ShengliJSAC09}\cite{ShengliJSAC09}. It maps $Y_R$ to $S_A \oplus S_B$. Note that the number of symbols in $Y_R$ is more than the number of symbols in $S_A \oplus S_B$ because of channel coding. Importantly, CNC involves both channel decoding and network coding, since CNC decodes the received signal $Y_R$ not to $S_A$ and $S_B$ individually, but to the network-coded source packet $S_A \oplus S_B$.

\emph{Part 2:}
The relay channel encodes $S_A  \oplus S_B $ to $X_R  = \Gamma(S_A  \oplus S_B )$. The relay then broadcast $\Gamma(S_A  \oplus S_B )$ to nodes $A$ and $B$.

Note that the channel coding in \emph{Part 2} is exactly the same as that in conventional channel-coded point-to-point communication link. Thus, the new distinct element introduced by PNC is the CNC process in \emph{Part 1}. As mentioned in \referred{ShengliJSAC09}\cite{ShengliJSAC09} and \referred{PNCSurvey}\cite{PNCSurvey}, the CNC component is unique to PNC, and different designs can have different performances and different implementation complexities. We refer interested readers to \referred{PNCSurvey}\cite{PNCSurvey} for a general discussion on different CNC designs. In this paper, we will study two specific CNC designs referred to as Jt-CNC and XOR-CD in Sections \ref{Sec:CPNC2} and \ref{Sec:CPNC3}, respectively.

\subsection{Jt-CNC: A Joint Channel-decoding and Network-Coding Scheme} \label{Sec:CPNC2}
In this subsection, we investigate the Jt-CNC scheme in which channel decoding and network coding are performed jointly in an integrated manner. Before presenting the scheme, we define asynchronous channel-coded PNC formally, and provide a quick overview of the RA code as the background material.



\begin{definition}[Asynchronous Channel-coded PNC]
In asynchronous channel-coded PNC, the two end nodes transmit their packets $X_A$ and $X_B$ with channel coding operation $\Gamma$ on their corresponding source packets $X_i$ (i.e., $X_i  = \Gamma \left( {S_i } \right)$ with $i \in \{ A,B\}$) in an asynchronous manner, so that the relay node $R$ receives the combined signals with $\phi  \ne 0$ and/or $\Delta  \ne 0$. The received baseband packet at $R$ is $Y_R  = X_A  + X_B  + W_R$ with $2N + 1$ symbols. Relay $R$ transforms $Y_R$ into a network-coded packet $X_R  = f(Y_R )$ with $N$ symbols for transmission in the downlink phase. Note that here we use $X_A$ to denote $\left( {x_A [1],x_A [1],\dots,x_A [N],x_A [N],0} \right)$ and $X_B$ to denote $\left( {0,x_B [1],x_B [1],\dots,x_B [N],x_B [N]} \right)$, respectively.
\end{definition}

\vspace{0.06in}
\subsubsection{Overview of RA code}

\begin{figure}[h]
\centering
\includegraphics[width=0.8\textwidth]{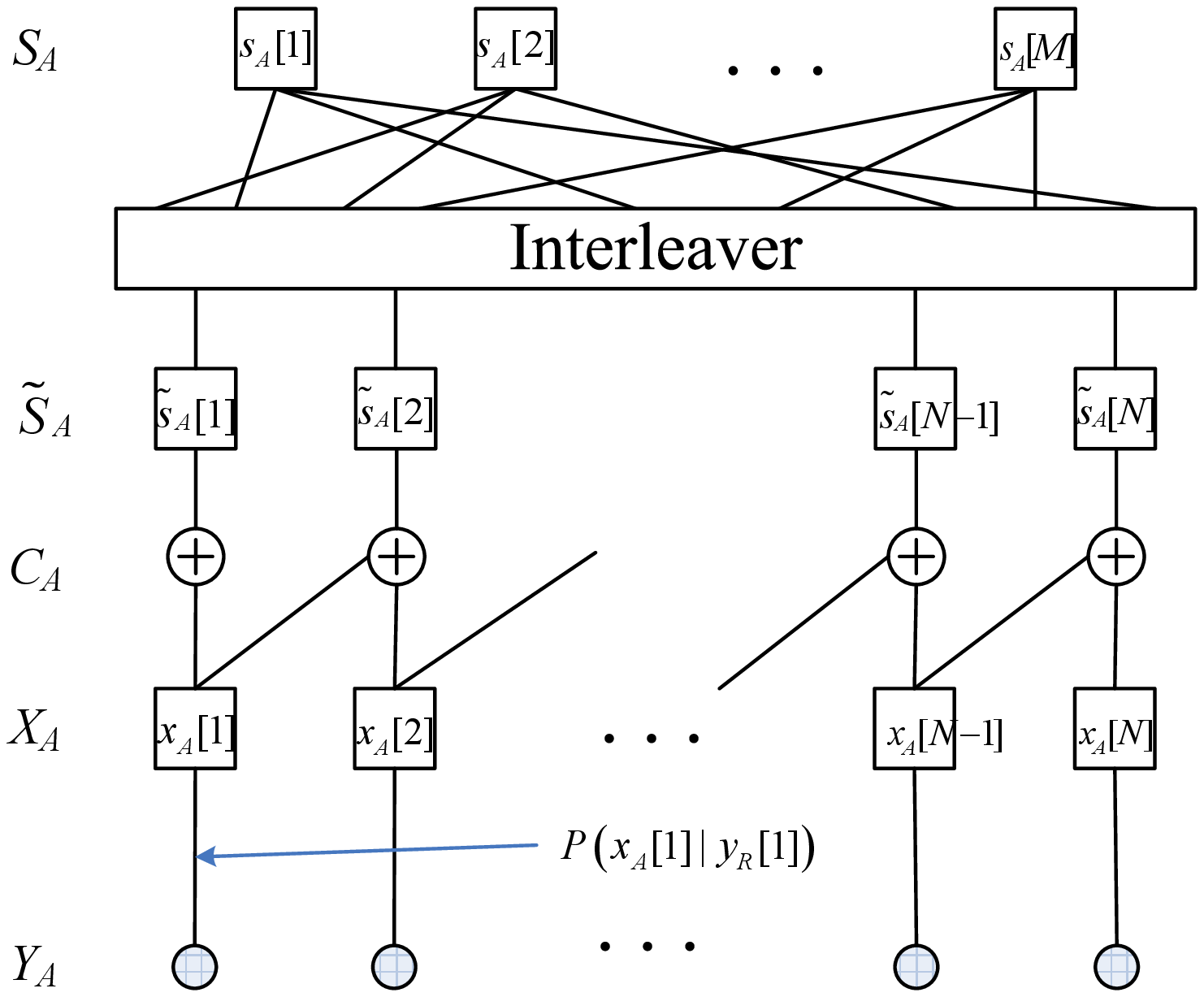}
\caption{Tanner graph for standard RA code.} \label{fig:RATannerGraph}
\end{figure}

Fig. \ref{fig:RATannerGraph} shows the encoding (decoding) Tanner graph \referred{YedTR01}\cite{YedTR01} of a standard RA code when used in a point-to-point communication link. The encoding process is as follows: from top to bottom in Fig. \ref{fig:RATannerGraph}, each source symbol in $S_A$ is first repeated $q$ times ($q = 3$ in Fig. \ref{fig:RATannerGraph}); then an interleaver is applied to decorrelate the adjacent repeated bits to get $\tilde{S}_A$; after that, the scrambled bits are accumulated by running XOR operations (represented by nodes $C_A$) to get the channel coded bits (nodes $X_A$) for transmission.

In PNC, the relay receives the channel-coded signals from nodes $A$ and $B$ simultaneously. We need to construct a modified Tanner graph for decoding purposes at the relay. We now present the Tanner graph of Jt-CNC. The Tanner graph of Jt-CNC is designed for the computation of $P(s_A [m],s_B [m]|Y_R )$ for $m = 1, \dots, M$. Once these probabilities are found, we can then obtain the PNC mapping by
\begin{align}
s_A [m] \oplus s_B [m] &= \arg \mathop {\max }\limits_{s}  P(s_A [m] \oplus s_B [m] = s|Y_R ) \nonumber \\
&= \arg \mathop {\max }\limits_{s} \sum\limits_{\scriptstyle (s_A [m],s_B [m]):\ \scriptstyle s_A [m] \oplus s_B [m] = s } {P(s_A [m], s_B [m]|Y_R )}. \label{equ:CpncDec}
\end{align}




\subsubsection{Tanner Graph of Jt-CNC and Message Update Rules}


As in the unchannel-coded case, we make use of the oversampled symbols in (\ref{equ:ApncSys4}) to construct a Tanner graph \referred{KschIT01}\cite{KschIT01} as shown in Fig. \ref{fig:TannerGraphAsyncCPNC}(b). In the Tanner graph, $Y_R$ is the evidence nodes, and there are $2N+1$ such nodes. For example, when RA code with repeat factor of 3 is used, $N = 3M$. In Fig. \ref{fig:TannerGraphAsyncCPNC}(b), $\Psi$ is the constraint nodes (also known as the compatibility nodes), and $S$ is the source nodes. Note that Fig. \ref{fig:TannerGraphAsyncCPNC}(b) is just the cascade of the Tanner graph in Fig. \ref{fig:TannerBPPNC} (without the PNC mapping represented by the triangles) and the Tanner graph in Fig. \ref{fig:RATannerGraph}, except that in Fig. \ref{fig:TannerGraphAsyncCPNC}(b) each source node $S$ (or code node $X$) is a pair that contains $(s_A[n], s_B[n])$ (or $(x_A[n], x_B[n])$).



When decoding, what is fed to the Tanner graph through the evidence nodes at the bottom are $P(x_A[n], x_B[n-1]|y_R[2n-1])$ and $P(x_A[n], x_B[n]|y_R[2n])$ for $n = 1, \dots, N$. This is illustrated in Fig. \ref{fig:TannerGraphAsyncCPNC}(a).

Let us explain the details assuming the use of QPSK modulation. With reference to (\ref{equ:ApncBPIni2}), we have $x_A [n] = a/\sqrt 2 $ and $x_B [n] = b/\sqrt 2$, where $a,b \in \{ 1 + j, - 1 + j, - 1 - j,1 - j\}$. Define $p_k^{a,b}  = P(x_A [\left\lceil {k/2} \right\rceil ] = {a \mathord{\left/ {\vphantom {a {\sqrt 2 }}} \right. \kern-\nulldelimiterspace} {\sqrt 2 }},x_B [\left\lfloor {k/2} \right\rfloor ] = {b \mathord{\left/ {\vphantom {b {\sqrt 2 }}} \right. \kern-\nulldelimiterspace} {\sqrt 2 }}|y_R [k])$,  which takes on values as in (\ref{equ:ApncBPIni1}) and (\ref{equ:ApncBPIni2}) in Section \ref{Sec:UPNC2}. Note that $p_k^{a,b}$ is computed based on $y_R [k]$ only.

With the above evidence node values and the Tanner graph structure, we then derive a set of message update rules for Jt-CNC. The detailed sum-product message-update rules for the messages associated with the edges in the Tanner graph can be found in in Appendix \ref{Sec:AppendB}. With reference to Fig. \ref{fig:TannerGraphAsyncCPNC}(b), in each iteration we update the messages in the following sequence: 1) right-bound messages below $X$; 2) left-bound messages below $X$; 3) upward-bound messages above $X$; 4) downward-bound messages above $X$. This sequence of message updates is repeated until the joint probabilities for the source nodes $P(s_A [m],s_B [m]|Y_R )$ $\forall m$ converge.

Compared with the Tanner graph for BP-UPNC in Fig. \ref{fig:TannerBPPNC}, the Tanner graph for Jt-CNC in Fig. \ref{fig:TannerGraphAsyncCPNC}(b) does not have a tree structure anymore. Thus, the joint probabilities ${P(s_A [m] \oplus s_B [m]|Y_R )}$ computed by BP are only approximations \referred{YedTR01}\cite{YedTR01}. Furthermore, multiple rounds of message updates for the same messages will be needed \referred{YedTR01}\cite{YedTR01}.

\subsection{XOR-CD: A Disjoint Channel-decoding and Network-Coding Scheme} \label{Sec:CPNC3}
We now look at the XOR-CD scheme in which channel decoding and network coding are performed in a disjoint manner. This scheme was studied in \referred{PopICC07}\cite{PopICC07}. It was also investigated in \referred{ZorziSPAWC09}\cite{ZorziSPAWC09} in the context of an OFDM-PNC system. Recently, it has been implemented in \referred{FPNC11}\cite{FPNC11} in the frequency domain via software radio.

Our goal here is to benchmark the performance of Jt-CNC with this scheme. Compared with Jt-CNC, this scheme (referred to as $\text{CNC}_\text{XOR-CD}$, or simply XOR-CD) is simpler to implement. The performance, however, is not as good.

\begin{figure}[h]
\centering
\includegraphics[width=0.9\textwidth]{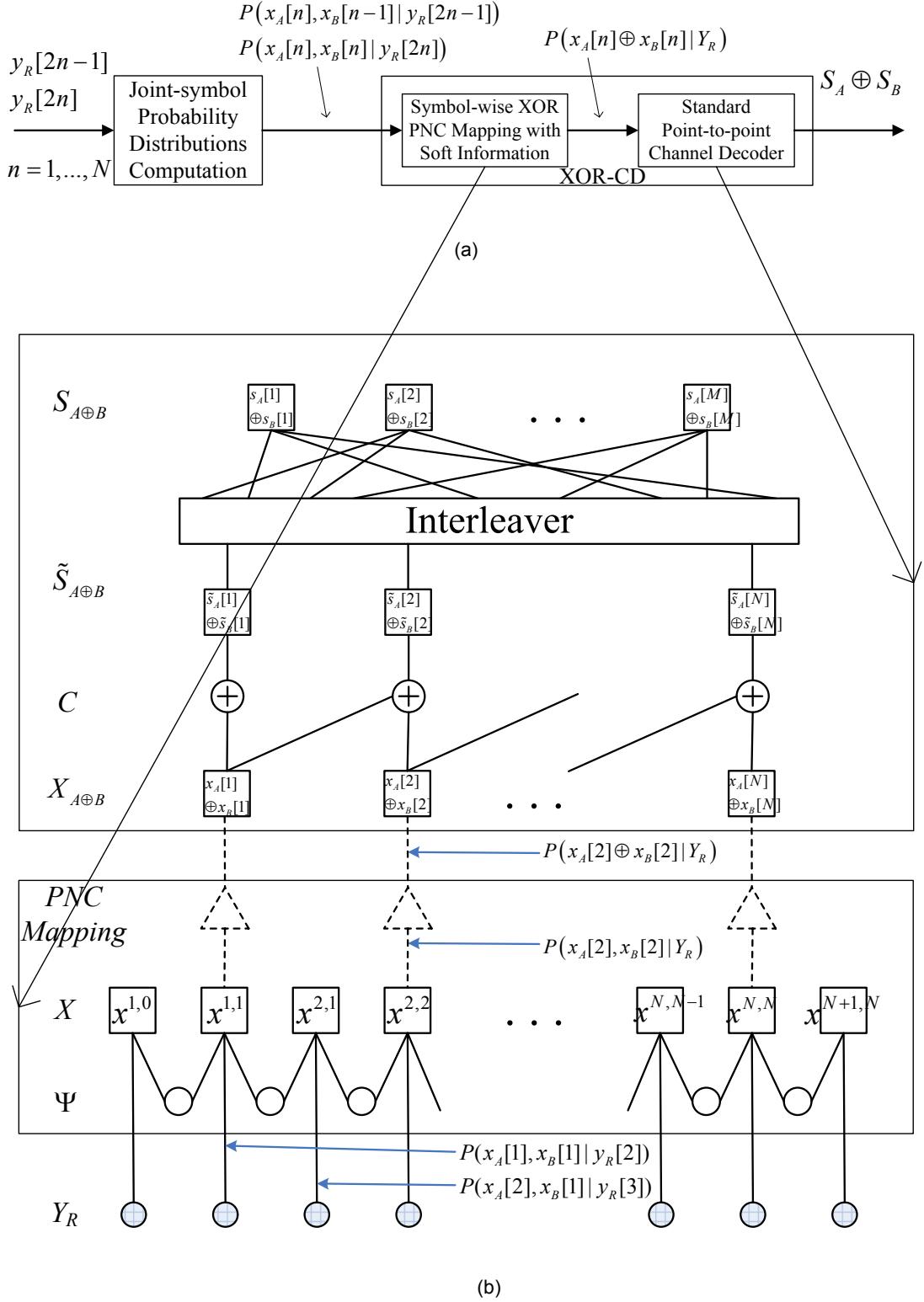}
\caption{Schematic diagram and Tanner graph for XOR-CD. (a) XOR-CD schematic diagram; (b) Tanner graph for XOR-CD in asynchronous channel-coded PNC. The triangle nodes perform the PNC mapping described in Section \ref{Sec:UPNC2}. Note that the channel decoder in the upper block is the standard RA decoder shown in Fig. \ref{fig:RATannerGraph}.} \label{fig:TannerGraphXORCD}
\end{figure}

Fig. \ref{fig:TannerGraphXORCD}(a) shows the schematic of XOR-CD. The acronym XOR-CD refers to a two-step process, in which we first apply the PNC mapping on the channel-coded symbols to obtain information on the XOR of the channel-coded symbols: $x_A[n] \oplus x_B [n], n = 1,\dots,N$; after that, we perform channel decoding on $X_A \oplus X_B$ to obtain $S_R  = S_A \oplus S_B $.

The first block in Fig. \ref{fig:TannerGraphXORCD}(a) computes soft information in the form of the probability distributions of XORed successive symbol pairs: $P(x_A[n] \oplus x_B[n]|Y_R)$ for $n = 1,\dots,N$. The computation performed by the first block is exactly the same as the PNC mapping in unchannel-coded PNC (see Section \ref{Sec:UPNC2}) except that now we apply the mapping on the channel-coded symbols. That is, the first block in Fig. \ref{fig:TannerGraphXORCD}(a) applies the BP-UPNC algorithm proposed in Section \ref{Sec:UPNC2}. This block does not exploit of the correlations among the successive symbols induced by channel coding.

The second block makes use of $P(x_A[n] \oplus x_B[n]|Y_R)$ from the first block to perform channel decoding to obtain the pairwise XOR of the source symbols, as shown in the upper block of Fig. \ref{fig:TannerGraphXORCD}(b). The channel decoder in the second block can be exactly the same as the channel decoder of a conventional point-to-point link if $\Gamma$ is a linear channel code (note: the RA code is linear), as implied by the following result: $ \Gamma(S_R) = \Gamma(S_A) \oplus \Gamma(S_B) = X_A  \oplus X_B  \Rightarrow S_R  = \Gamma^{ - 1} (X_A  \oplus X_B )$. This is the reason why XOR-CD is simpler to implement than Jt-CNC\footnote{The complexity of Jt-CNC under QPSK modulation is due to the 16 combinations of $(x_A[n],x_B[n])$ and $(s_A[n], s_B[n])$ in the branches of the Tanner graph that the sum-product algorithm has to compute over (see Fig. \ref{fig:TannerGraphAsyncCPNC}(b)). For XOR-CD, each branch has only 4 combinations in the channel decoding part, thanks to the XOR operation prior to channel decoding (see Fig. \ref{fig:TannerGraphXORCD}(b)).}. However, the first block loses some useful information in only presenting $P(x_A[n] \oplus x_B[n]|Y_R)$, $n=1,\dots,N$, to the second block rather then the joint probabilities $P(x_A[n], x_B[n-1]|y_R[2n-1])$ and $P(x_A[n], x_B[n]|y_R[2n])$, $n=1,\dots,N$.

\subsection{Numerical Results}\label{Sec:CPNC4}

This subsection presents simulation results of Jt-CNC and XOR-CD.


\subsubsection{Summary of Results}
Our simulations yield the following findings:
\begin{itemize}
\item In Jt-CNC, rather than having BER performance penalty due to phase asynchrony (as in BP-UPNC), we have phase reward.
\item In Jt-CNC, the performance spread of QPSK channel-coded PNC due to different phase offsets is small (no more than 1 dB), with or without symbol misalignment.
\item Jt-CNC achieves significantly better BER performance compared with XOR-CD (3 dB on average) for both BPSK and QPSK.
\item In Jt-CNC, as in BP-UPNC, symbol misalignment also desensitizes the system performance to phase asynchrony.
\end{itemize}

The general conclusion is that in channel-coded PNC, phase and symbol asynchronies are not performance limiting factor with an appropriate CNC algorithm, such as Jt-CNC.
\subsubsection{Detailed Description}

\paragraph{Jt-CNC}

\begin{figure}[h]
\centering
\includegraphics[width=1\textwidth]{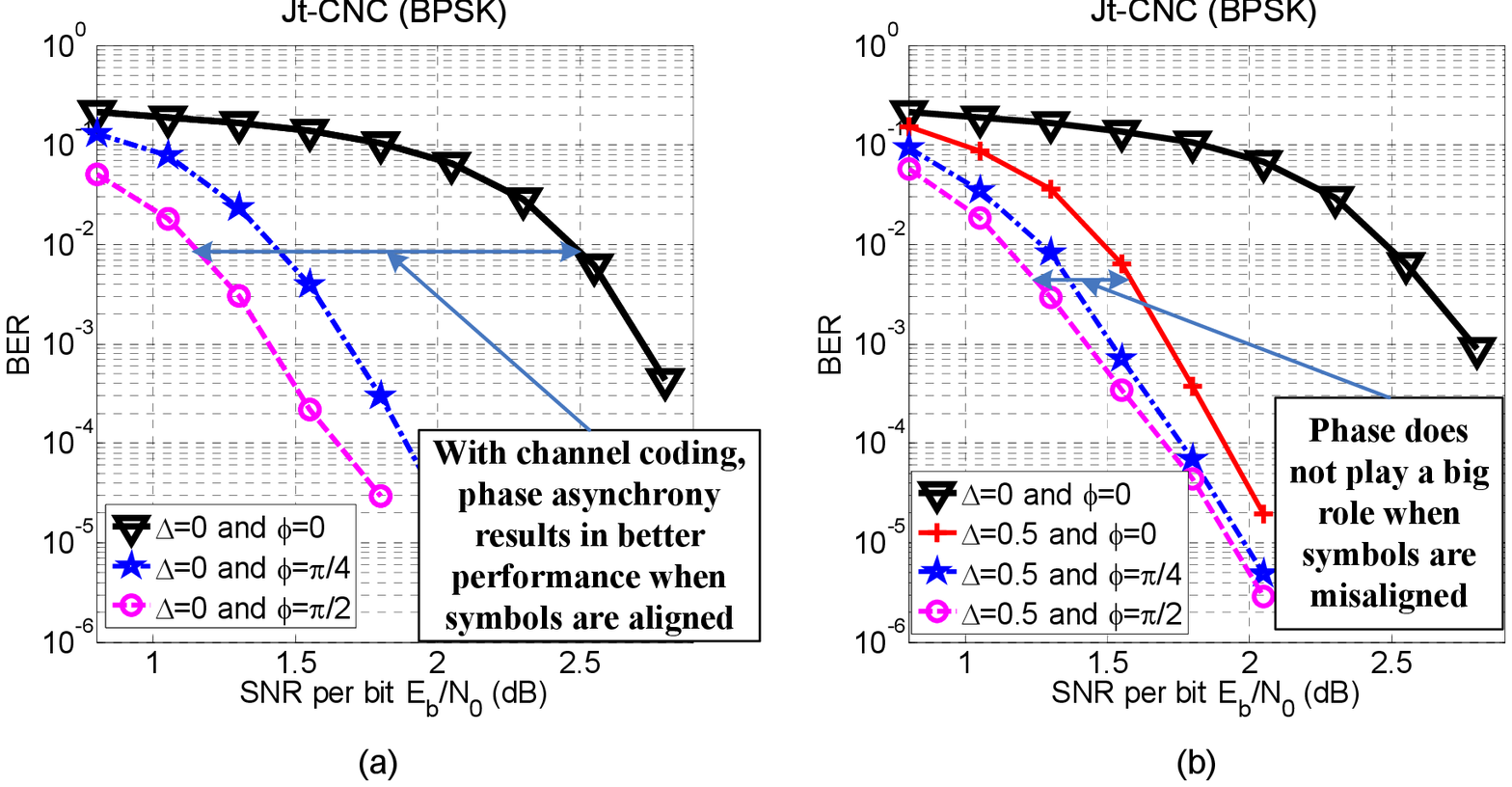}
\caption{BER of the uplink XORed value $s_A[m] \oplus s_B[m]$ in Jt-CNC for BPSK modulated channel-coded PNC using RA code with repeat factor three : (a) Jt-CNC without symbol asynchrony ($\Delta = 0$) ; (b) Jt-CNC with symbol asynchrony ($\Delta \ne 0$). Note that $E_b$ is energy per source bit here.}
\label{fig:BPCPNCsimulationBPSK}
\end{figure}

\begin{figure}[h]
\centering
\includegraphics[width=1\textwidth]{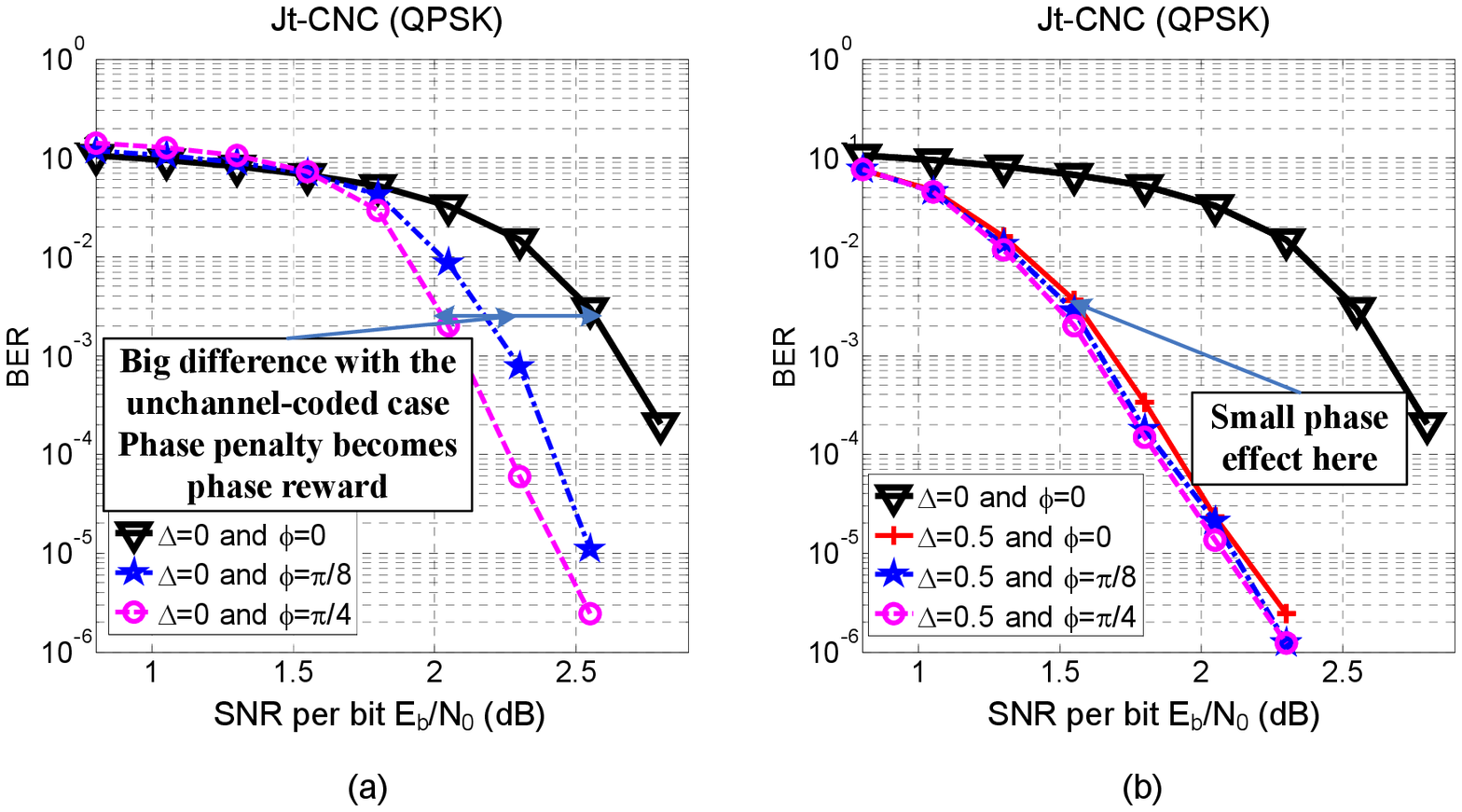}
\caption{BER of the uplink XORed value $s_A[m] \oplus s_B[m]$ in Jt-CNC for QPSK modulated channel-coded PNC using RA code with repeat factor three : (a) Jt-CNC without symbol asynchrony ($\Delta = 0$) ; (b) Jt-CNC with symbol asynchrony ($\Delta \ne 0$). Note that $E_b$ is energy per source bit here.}
\label{fig:BPCPNCsimulationQPSK}
\end{figure}

We adopt the regular RA code with a coding rate of 1/3 in our simulations. Fig. \ref{fig:BPCPNCsimulationBPSK} and Fig. \ref{fig:BPCPNCsimulationQPSK} show the simulation results of BPSK and QPSK modulated Jt-CNC schemes, respectively. The $x$-axis is the average per-bit SNR. For fair comparison between unchannel-coded PNC and channel-coded PNC, we shift the curves of the latter by $10\log _{10}3$ dB to the right to take into account that each bit is repeated 3 times in our RA channel coding.

For each data point of BPSK, we simulated 10,000 packets of 2,048 bits. We benchmark the results against the perfect synchronous case where $\Delta =0$ and $\phi = 0$. For each data point of QPSK, we simulated 10,000 packets of 4,096 bits. These 4,096 bits are divided into in-phase and quadrature parts, each having 2,048 bits. The total numbers of inputs (symbols) in the Tanner graph of Fig. \ref{fig:TannerGraphAsyncCPNC}(b) are the same for both the BPSK and QPSK cases.

We see from Fig. \ref{fig:BPCPNCsimulationBPSK} and Fig. \ref{fig:BPCPNCsimulationQPSK} that instead of penalty, phase asynchrony actually improves the performance in channel-coded PNC. In the case of QPSK unchannel-coded PNC, recall from Fig. \ref{fig:BPPNCsimulationQPSK}(a) that when symbols are aligned, the penalty can be as high as 6-7 dB. However, with channel coding, as can be seen from Fig. \ref{fig:BPCPNCsimulationQPSK}(a), the penalty goes away and become a reward of around 0.5 dB. In the case of symbol misalignment of 0.5 symbol, the reward is around 1 dB.

Another interesting observation is that in the case of symbol misalignment, say $\Delta = 0.5$, the BER performance is less dependent on the degree of phase asynchrony. In particular, for QPSK, as can be seen from Fig. \ref{fig:BPCPNCsimulationQPSK}(b), the SNR required to achieve a target BER does not vary much with phase offset.

\begin{figure}[h]
\centering
\includegraphics[width=1\textwidth]{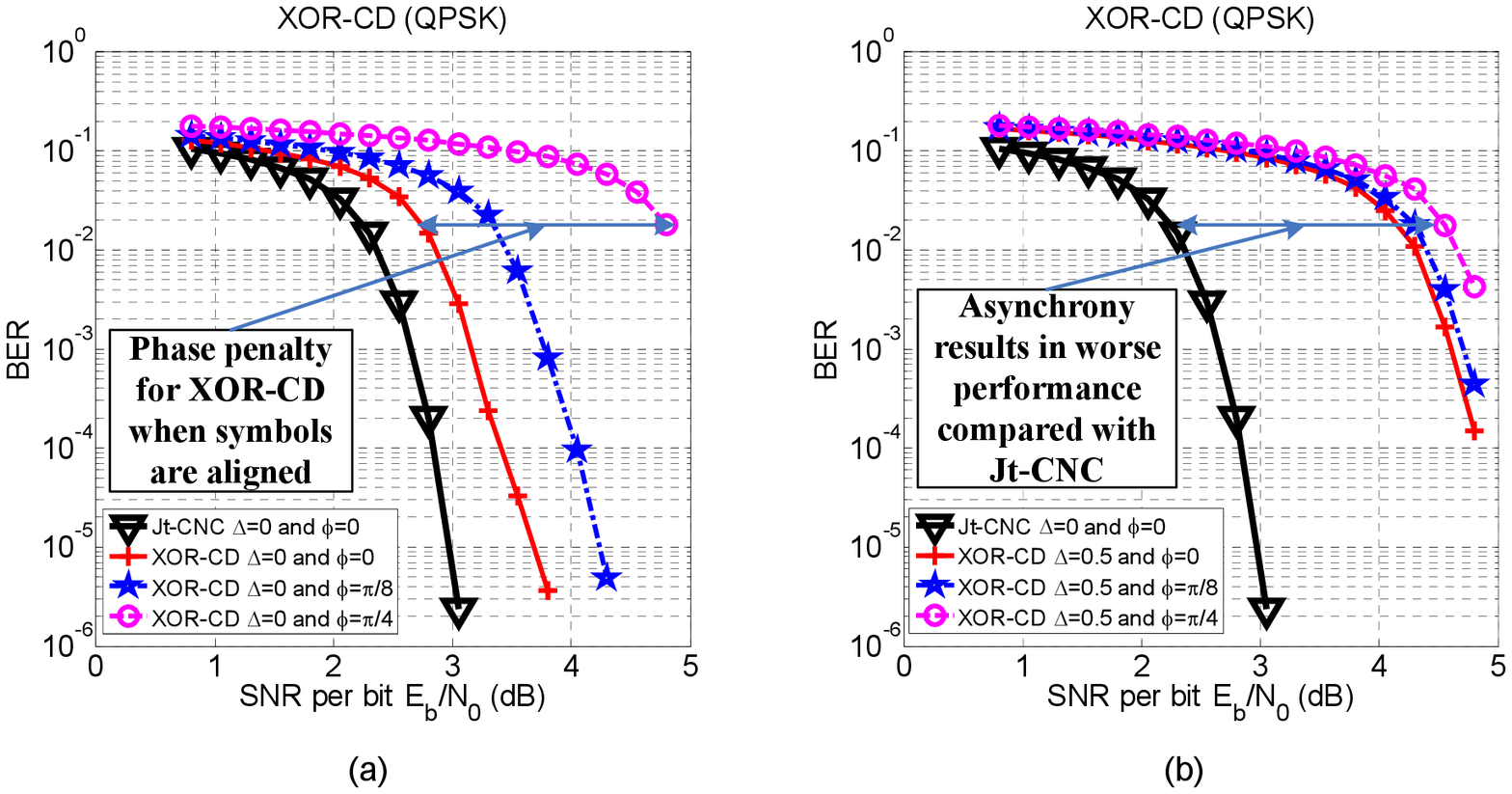}
\caption{BER of the uplink XORed value $s_A[m] \oplus s_B[m]$ in XOR-CD for QPSK modulated channel-coded PNC using RA code with repeat factor three : (a) XOR-CD without symbol asynchrony ($\Delta = 0$) ; (b) XOR-CD with symbol asynchrony ($\Delta \ne 0$). Note that $E_b$ is energy per source bit here.}
\label{fig:BerXORCDQPSK}
\end{figure}

\paragraph{XOR-CD}
Let us now look at the performance of XOR-CD. For asynchronous XOR-CD. Fig. \ref{fig:BerXORCDQPSK} shows the BER results of asynchronous XOR-CD with QPSK modulation. The simulation condition and parameters for XOR-CD are the same as that in Jt-CNC (i.e., for each data point of QPSK, we simulated 10,000 packets of 4,096 bits). In general, we can see that this scheme, although less complex than Jt-CNC, has significantly worse performance. In addition, instead of phase reward, there is phase penalty. Its performance is far from what could be achieved fundamentally.

%


\subsection{Diversity and Certainty Propagation in Jt-CNC}\label{CertProp1}

In Section \ref{CertProp}, we explained that for unchannel-coded PNC, symbol misalignment induces ``diversity and certainty propagation'' effects that improves the BER performance and makes the system robust against phase asynchrony. Similar diversity and certainty propagation effects are also present in channel-coded PNC when Jt-CNC is used. Unlike in unchannel-coded PNC, in which such effects occur only when symbols are misaligned, these positive effects are present in Jt-CNC with or without symbol misalignment. This is because when channel coding is used, the information on each source symbol is embedded in multiple channel-coded symbols regardless of symbol alignment. The negative effect of a ``bad'' channel-coded symbol pair may be overcome by the positive effect of a ``good'' channel-coded symbol pair during the channel-decoding process. Consequently, the issue of phase asynchrony becomes less critical in Jt-CNC. In particular, even when symbols are perfectly aligned, channel-coded PNC with Jt-CNC can remove the phase penalty via diversity and certainty propagation.

For channel-coded PNC with XOR-CD, from Fig. \ref{fig:BerXORCDQPSK} we see that phase asynchrony still imposes significant penalties. Unlike in Jt-CNC, the certainty propagation effect is weakened in XOR-CD. This is because XOR-CD makes use of $P(x_A[n] \oplus x_B[n]|Y_R)$ rather than the joint probabilities $P(x_A[n], x_B[n-1]|y_R[2n-1])$ and $P(x_A[n], x_B[n]|y_R[2n])$ in the channel decoding process. With only $P(x_A[n] \oplus x_B[n]|Y_R)$, certainties in individual symbol $x_A[n]$ or $x_B[n]$ (for ``good'' symbol pairs) are lost and cannot be propagated (see Section \ref{CertProp} on the mechanism of certainty propagation, which requires certainties in individual symbols $x_A[n]$ or $x_B[n]$).

\section{Conclusions}\label{Sec:Conclusion}

This paper has investigated the effects of asynchronies in unchannel-coded and channel-coded PNC systems. Our investigations focus on the effects of carrier-phase and symbol-misalignment asynchronies. We use belief propagation (BP) algorithms at the relay to perform the network coding and, in the case of channel-coded PNC, channel decoding operations.

For unchannel-coded PNC, our BP algorithm is an exact maximum likelihood (ML) decoding algorithm to find symbol-wise XOR of the packets from the two end nodes. With our BP ML algorithm, we find that symbol misalignment has an advantage. In particular, with symbol asynchrony, the performance penalty due to phase asynchrony can be reduced drastically, thanks to some diversity and certainty propagation effects in our algorithm. Essentially, the BER performance is not very sensitive to different carrier-phase offsets anymore (around 0.5 dB spread for different phase offsets when the symbols are misaligned by half symbol, as opposed to more than 6 dB when the symbols are perfectly aligned). Our results suggest that if we could control the symbol offset, it would actually be advantageous to deliberately introduce symbol misalignment so that the system is robust against phase offsets.

For channel-coded PNC, we study two schemes, Jt-CNC and XOR-CD. In Jt-CNC, network coding and channel decoding at the relay are performed jointly in an integrated manner. In XOR-CD, network coding is first performed on the channel-coded symbols before channel decoding is applied; that is, the two processes are disjoint. Both Jt-CNC and XOR-CD make use BP, but in different ways.

Due to its better performance, we believe Jt-CNC is closer to what can ultimately be achieved in the channel-coded PNC system. In particular, we believe that the fundamental performance impacts of phase and symbol asynchronies are better revealed through the performance results of Jt-CNC.  We observe an interesting result that instead of performance penalties, symbol and phase asynchronies can actually give rise to performance rewards. In particular, in channel-coded PNC, unlike in unchannel-coded PNC, phase asynchrony is not a performance limiting factor even when symbols are perfectly aligned. Furthermore, the performance spread arising from all combinations of symbol and phase offsets is only slightly more than 1 dB. The intuitive explanation is as follows. With channel-coding, information on each source symbol is encoded into several channel-coded symbols. Jt-CNC makes use of the correlations among different channel-coded symbols to achieve diversity and certainty propagation effects akin to those in symbol-misaligned unchannel-coded PNC.

Prior to this work, it has often been thought that strict synchronization is needed for PNC. Our work suggests, however, the penalties due to asynchronies can be nullified to a large extent. In unchannel-coded PNC, symbol asynchrony can reduce the negative effects of phase asynchrony significantly. In channel-coded PNC, with the use of Jt-CNC, both phase and symbol asynchronies have small effects on the performance; and these are positive effects.

\bibliographystyle{IEEEtran}
\bibliography{apnc}

\appendices

\section{Message Update Steps of Jt-CNC}\label{Sec:AppendB}
\noindent\emph{Step 1. Updates of messages below code nodes $X$:}

\begin{figure}[h]
\centering
\includegraphics[width=0.9\textwidth]{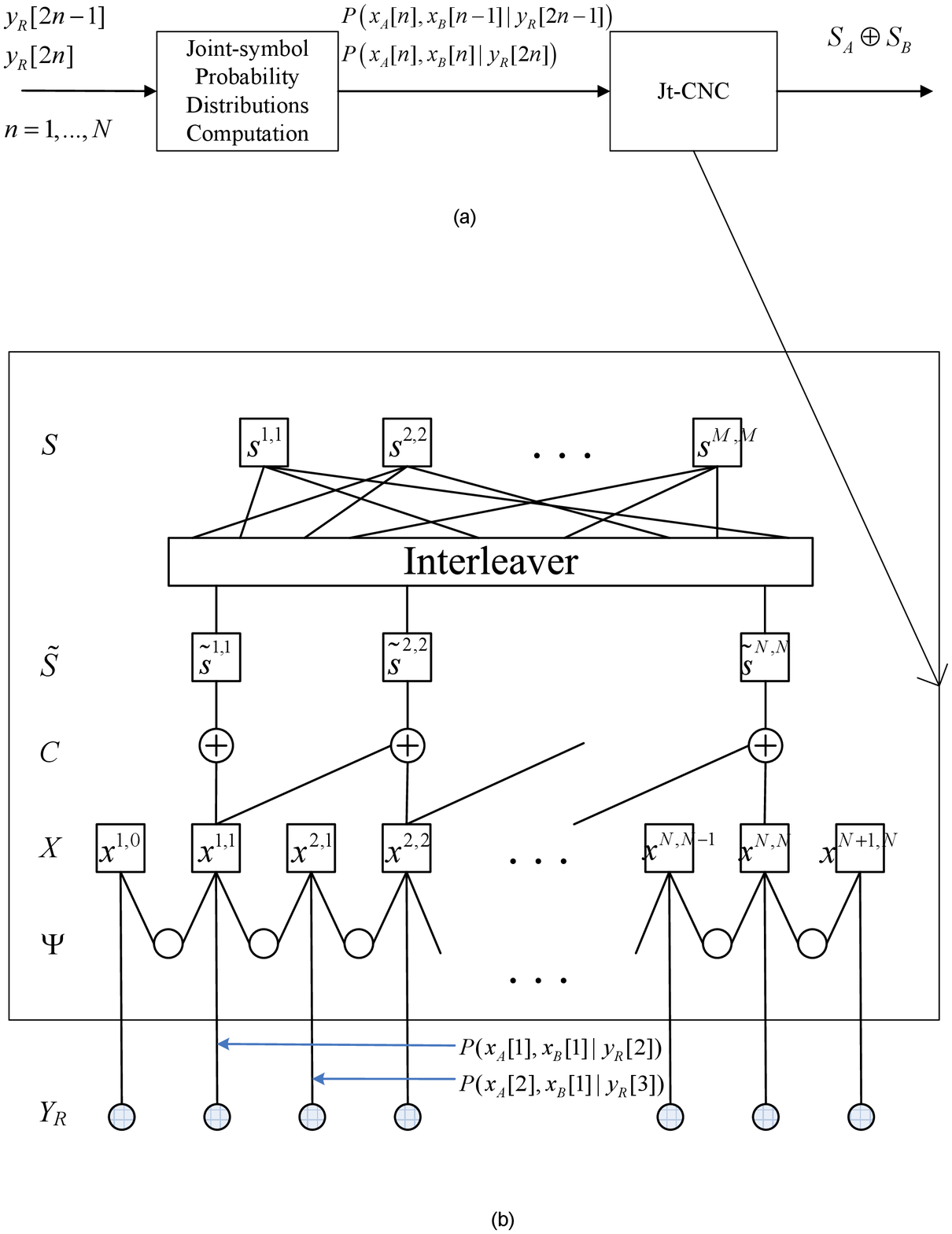}
\caption{Schematic diagram and Tanner graph for Jt-CNC. (a) Jt-CNC schematic diagram; (b) Tanner graph for Jt-CNC in asynchronous channel-coded PNC. $s^{m,m}$ stands for $\left( {s_A [m],s_B [m]} \right)$, $x^{n,n}$ stands for $\left( {x_A [n],x_B [n]} \right)$, and $\tilde{s }^{n,n}$ stands for $ \left( {\tilde{s}_A[n],\tilde{s}_B[n]} \right)$, which are the repeated symbols after the interleaver.} \label{fig:TannerGraphAsyncCPNC}
\end{figure}

\begin{figure}[h]
\centering
\includegraphics[width=1\textwidth]{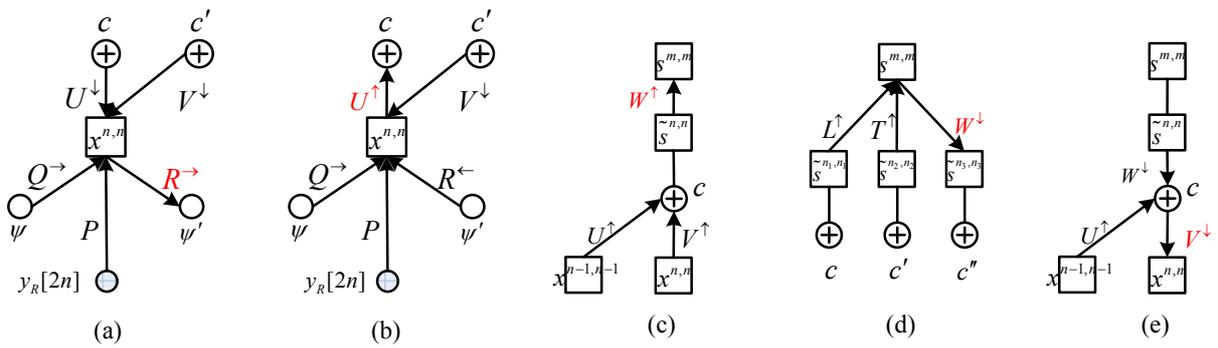}
\caption{Jt-CNC message update procedure. (a) Update of a right-bound message below code node $x^{n,n}$; (b) Update of an upward message into check node $c$; (c) Update of upward messages into a source node $s^{m,m}$; (d) Update of a downward message into a check node $c''$; (e) Update of a downward message into a code node $x^{n,n}$.} \label{fig:MessageUpdate}
\end{figure}

We consider the update of a right-bound message below $X$ in Fig. \ref{fig:TannerGraphAsyncCPNC}(b). The update of a left-bound message below $X$ is similar. For the message emanating out of an odd node $x^{n + 1,n}$, the structure of the Tanner graph is the same to that in Section \ref{Sec:UPNC2}. Thus, the update rule for an odd node $x^{n + 1,n}$ is the same and is omitted here. Similarly, messages emanating out of a compatibility node $\Psi$ are updated the same way as in Section \ref{Sec:UPNC2}. For the message emanating out of an even node, $x^{n,n}$, with reference to Fig. \ref{fig:MessageUpdate}(a), unlike the structure in Section \ref{Sec:UPNC2}, there are two additional incoming messages, $U^\downarrow$ and $V^\downarrow$, coming from two check nodes from above. The messages $U^\downarrow$ and $V^\downarrow$ have to be taken into account when updating $R^\to = (r^{1 + j,1 + j} ,r^{1 + j, - 1 + j} ,\dots,r^{1 - j,1 - j})$. The update of $R^\to$ is based on the $Q^\to = (q^{1 + j,1 + j},q^{1 + j, - 1 + j},\dots,q^{1 - j,1 - j})$, $U^\downarrow   = (u^{1 + j,1 + j}, u^{1 + j, - 1 + j},\dots,u^{1 - j,1 - j})$, $V^\downarrow = (v^{1 + j,1 + j}, v^{1 + j, - 1 + j},\dots,v^{1 - j,1 - j} )$, and $P = (p^{1 + j,1 + j} ,p^{1 + j, - 1 + j} ,\dots,p^{1 - j,1 - j} )$. Specifically,
\begin{align}
r^{a,b} = \beta _1 p^{a,b} q^{a,b} u^{a,b} v^{a,b}, \label{equ:AcpncBPUpdate1}
\end{align}
where $a,b \in \chi$, where $\chi$ was defined in Section \ref{Sec:UPNC2}1 as the alphabet of $\{ 1 + j, - 1 + j, - 1 - j,1 - j \}$ and $\beta_1$ is the normalization factor to ensure that the sum of probabilities, $\sum\limits_{a,b \in \chi} {r^{a,b}} = 1$.

\vspace*{.04in}
\noindent\emph{Step 2. Updates of upward messages into check nodes $C$:}

With reference to Fig. \ref{fig:MessageUpdate}(b), we consider the update of an upward message $U^\uparrow$ into check node $c$. The update of the upward message $V^\uparrow$ into check node $c'$ is similar. Note that only even code nodes $x^{n,n}$ are associated with the update of messages into check nodes, and we need not consider $x^{n+1,n}$. With reference to Fig. \ref{fig:MessageUpdate}(b), the information from $\psi$ and $\psi'$ is passed by the two additional messages $Q^\to$ and $R^\leftarrow$. With the notations $U^\uparrow = (u^{1 + j,1 + j}, u^{1 + j, - 1 + j},\dots,u^{1 - j,1 - j})$, $Q^\to = (q^{1 + j,1 + j}, q^{1 + j, - 1 + j},\dots,q^{1 - j,1 - j})$, $R^\leftarrow = (r^{1 + j,1 + j}, r^{1 + j, - 1 + j},\dots,r^{1 - j,1 - j})$, $V^\downarrow = (v^{1 + j,1 + j}, v^{1 + j, - 1 + j},\dots,v^{1 - j,1 - j})$ and $P = (p^{1 + j,1 + j},\\ p^{1 + j, - 1 + j},\dots,p^{1 - j,1 - j})$, the update rule is given by
\begin{align}
u^{a,b}  = \beta _2 p^{a,b} q^{a,b} r^{a,b} v^{a,b}, \label{equ:AcpncBPUpdate2}
\end{align}
where $a,b \in \chi$ and $\beta_2$ is the normalization factor to ensure $\sum\limits_{a,b \in \chi} {u^{a,b}} = 1$.

\vspace*{.04in}
\noindent\emph{Step 3. Update of upward messages into the source nodes $S$:}

We consider the update of an upward message into source node $s^{m,m}$. With reference to Fig. \ref{fig:MessageUpdate}(c), we derive the update equations for the upward message $W^\uparrow = (w^{1 + j,1 + j}, w^{1 + j, - 1 + j}, \dots, \\ w^{1 - j,1 - j} )$ from a check node $c$ to a source node $s^{m,m}$. For QPSK, each input to the XOR check node has 16 possibilities. With two inputs $U^\uparrow = (u^{1 + j,1 + j}, u^{1 + j, - 1 + j},\dots,u^{1 - j,1 - j})$ and $V^\uparrow = (v^{1 + j,1 + j}, v^{1 + j, - 1 + j},\dots,v^{1 - j,1 - j})$, there are 256 possible input combinations, and these 256 possibilities will be mapped to 16 possibilities in the output (i.e., each $w^{a,b}$ is a summation of 16 input probabilities.). Based on the sum-product principle of the BP algorithm, we have
\begin{align}
w^{a,b} = \sum\limits_{\scriptstyle a_u  \oplus a_v  = a \hfill \atop \scriptstyle b_u  \oplus b_v  = b \hfill} {u^{a_u ,b_u } v^{a_v ,b_v } }, \label{equ:AcpncBPUpdate3}
\end{align}
where $a,b,a_u ,b_u ,a_v ,b_v  \in \chi$ Note that here, the QPSK XOR $a = a_u \oplus a_v$ means the combination of its corresponding XORed real and imaginary parts, $a = a_u^{{\mathop{\rm Re}\nolimits} }  \cdot a_v^{{\mathop{\rm Re}\nolimits} }  + ja_u^{{\mathop{\rm Im}\nolimits} }  \cdot a_v^{{\mathop{\rm Im}\nolimits} }$, where $a_u  = a_u^{{\mathop{\rm Re}\nolimits} }  + ja_u^{{\mathop{\rm Im}\nolimits} }$ and $a_v = a_v^{{\mathop{\rm Re}\nolimits} } + ja_v^{{\mathop{\rm Im}\nolimits}}$ (e.g., if $\left( {a_u ,a_v } \right) = \left( { - 1 + j,1 + j} \right)$ then $a = \left( { - 1 \cdot 1} \right) + j\left( {1 \cdot 1} \right) =  - 1 + j$).

\vspace*{.04in}
\noindent\emph{Step 4. Update of downward messages into the check nodes $C$:}

We next proceed to the updates of downward messages flowing top to bottom. With reference to Fig. \ref{fig:MessageUpdate}(d), we derive the message update rules for a downward message $W^\downarrow = (w^{1 + j,1 + j}, w^{1 + j, - 1 + j},\dots,w^{1 - j,1 - j})$ from a source node $s^{m,m}$ to a check node $c''$, based on the upward messages $L^\uparrow = (l^{1 + j,1 + j}, l^{1 + j, - 1 + j},\dots,l^{1 - j,1 - j})$ and $T^\uparrow = (t^{1 + j,1 + j}, t^{1 + j, - 1 + j},\dots,\\l^{1 - j,1 - j})$. Again, based on the compatible sum-product principle of BP, we have
\begin{align}
w_{}^{a,b}  = \beta _3 l_{}^{a,b} t_{}^{a,b}, \label{equ:AcpncBPUpdate4}
\end{align}
where $a,b \in \chi$ and $\beta _3$ is the normalization factor to ensure that the sum of probabilities, $\sum\limits_{a,b \in \chi} {w_{}^{a,b} } = 1$.

\vspace*{.04in}
\noindent\emph{Step 5. Updates of downward messages into code nodes $X$:}

With reference to Fig. \ref{fig:MessageUpdate}(e), the update equations for the message from a check node to a code node are similar to those from a check node to a source node in Fig. \ref{fig:MessageUpdate}(c), except for the update direction. So we omit the details here.

\pagebreak


\begin{table}[h]
    \caption{Four cases of PNC systems}
    \centering
    \begin{tabular}{|c|c|c|}
    \hline \multicolumn{1}{|c|}{} &\multicolumn{1}{|c|}{$\Delta  = 0$} &\multicolumn{1}{|c|}{$\Delta  \ne 0$}
    \\\hline $\phi  = 0$ & $\phi  = 0,\Delta  = 0$ (Case 1) & $\phi  = 0,\Delta  \ne 0$ (Case 2)
    \\\hline $\phi  \ne 0$ & $\phi  \ne 0,\Delta  = 0$ (Case 3) & $\phi \ne 0,\Delta  \ne 0$ (Case 4)
    \\\hline
    \end{tabular}
    \tabcolsep -10mm\label{tab:Class}
\end{table}
\vspace{-.2in}
    \centering\scriptsize $\Delta$ is the symbol offset and $\phi$ is the phase offset

\vspace{.8in}

\pagebreak

\end{document}